\documentclass[proof]{pasj00}
\draft 
\begin{document}


\SetRunningHead{T. Goto et al.}{The Environment of Passive Spiral Galaxies in the SDSS}
\Received{2003/01/16}
\Accepted{2003/06/23}
\date{\today}
\title{The Environment of Passive Spiral Galaxies in the SDSS}

\author{%
   Tomotsugu \textsc{Goto},\altaffilmark{1,2}
   Sadanori \textsc{Okamura},\altaffilmark{2}
   Maki \textsc{Sekiguchi},\altaffilmark{1}\\
   Mariangela \textsc{Bernardi},\altaffilmark{3}
   Jon \textsc{Brinkmann},\altaffilmark{4}
   Percy L. \textsc{G{\' o}mez},\altaffilmark{3}
   Michael  \textsc{Harvanek},\altaffilmark{4}\\ 
   Scot J.  \textsc{Kleinman},\altaffilmark{4} 
   Jurek  \textsc{Krzesinski},\altaffilmark{4,5} 
   Dan  \textsc{Long},\altaffilmark{4} 
   Jon  \textsc{Loveday},\altaffilmark{6} \\
   Christopher J. \textsc{Miller},\altaffilmark{3}
   Eric H.  \textsc{Neilsen},\altaffilmark{4}
   Peter R.  \textsc{Newman},\altaffilmark{4}  
   Atsuko  \textsc{Nitta},\altaffilmark{4} \\
   Ravi K. \textsc{Sheth},\altaffilmark{7}
   Stephanie A.  \textsc{Snedden},\altaffilmark{4}
   and Chisato \textsc{Yamauchi}\altaffilmark{8}
}   
 \altaffiltext{1}{Institute for Cosmic Ray Research, The University of
   Tokyo,\\ Kashiwanoha, Kashiwa, Chiba 277-0882
 \\E-mail (T.G.): yohnis@icrr.u-tokyo.ac.jp}
 \altaffiltext{2}{Department of Astronomy and Research Center for the
   Early Universe,\\ School of Science, The University of Tokyo, Tokyo
   113-0033}
 \altaffiltext{3}{Department of Physics, Carnegie Mellon University, \\
 5000 Forbes Avenue, Pittsburgh, PA 15213-3890, USA}
 \altaffiltext{4}{Apache Point Observatory,\\
     2001 Apache Point Road,
     P.O. Box 59, Sunspot, NM 88349-0059, USA}
\altaffiltext{5}{
Mt. Suhora Observatory, Cracow Pedagogical University, ul. Podchorazych 2, 30-084 Cracow, Poland
}
\altaffiltext{6}{
Sussex Astronomy Centre,
University of Sussex,
Falmer, Brighton BN1 9QJ, UK}
 \altaffiltext{7}{ Department of Physics and Astronomy,
 University of Pittsburgh\\
 3941 O'Hara Street
 Pittsburgh, PA 15260  }
\altaffiltext{8}{
National Astronomical Observatory, 2-21-1 Osawa, Mitaka, Tokyo 181-8588}


\maketitle

\begin{abstract}\label{abstract}
\date{\today}

 In previous work on galaxy clusters, several authors reported the
 discovery of an unusual 
 population of galaxies, which have spiral morphologies, but do not show
 any star-formation activity.  These galaxies are called ``passive
 spirals'', and have been interesting since it
 has been difficult to understand the existence of such galaxies. 
  Using a volume-limited sample (0.05$<z<$0.1 and $Mr^*<-20.5$; 25813
 galaxies) of the 
 Sloan Digital Sky Survey data, we found 73 (0.28$\pm$0.03\%)
 passive spiral galaxies and studied their environments.
 It is found that passive
 spiral galaxies exist in a local galaxy density of 1--2 Mpc$^{-2}$ and
 have a 1--10 cluster-centric virial radius. Thus, the origins of passive
 spiral galaxies 
 are likely to be cluster-related. These characteristic environments
 coincide with a previously reported environment where the galaxy
 star-formation rate suddenly declines and the so-called 
 morphology-density relation turns.
 It is likely that the same physical mechanism is
 responsible for all of these observational results.
  The existence of passive spiral
 galaxies suggests that a physical mechanism that works calmly is
 preferred to dynamical origins such as major merger/interaction since such a
 mechanism would destroy the spiral-arm structures. 
  Compared with the observed cluster galaxy evolution such as the
 Butcher--Oemler effect and the morphological Butcher--Oemler effect,
 passive spiral galaxies are likely to be a key galaxy population in
 transition between red, elliptical/S0 galaxies in low-redshift clusters
 and blue, spiral galaxies more numerous in higher-redshift clusters.

\end{abstract}
\date{\today}

\section{Introduction}\label{intro}

 Recent morphological studies of distant cluster galaxies have revealed the
 presence of an unusual population of galaxies with a spiral morphology
 and lack of star-formation activity (Couch et al. 1998; Dressler et
 al. 1999; Poggianti et al. 1999). The origins of these ``passive
 spirals'' have remained a mystery since it
 has been difficult to understand the existence of such galaxies.  
  The phenomena suggest that star formation in these system
 has ended calmly, without disturbing their spiral-arm structures. 
 Many people speculated that cluster-related phenomena might be
 responsible for creation of passive spiral galaxies since they are found
 during cluster studies.  However, it has not been well established
 whether 
 these phenomena are more relevant in clusters, or are common in the
 field regions as well, simply because it
 has been difficult to study this rare class of galaxies in the field
 region.    

    Also, the existence of a
  similar type of galaxies has been reported. Galaxies with a low
  arm/inter-arm contrast 
  in their disks were classified as anemic by  van den Bergh
  (1976). He found an excess of anemic spiral galaxies in the Virgo cluster.
  Various H{\scshape i} follow-up
  observations have revealed a lower gas density in anemic spiral galaxies,
  presumably lowering the star-formation rate and making the spiral arms
  smoother (Bothun, Sullivan 1980; Wilkerson 1980; Phillipps 1988;
  Cayatte et al. 1994;    Bravo-Alfaro et al. 2001).  
  Especially, Elmegreen et  al. (2002) found that the gas surface
  density of anemic 
  spirals is below the
  threshold for star formation (Kennicutt 1989), revealing that low star
  formation in anemic spirals in fact comes from a low gas density.
  Although the
  definition of anemic spiral galaxies is somewhat different from that
  of passive spirals, considering 
  similarities in properties (presence of spiral arms and lack of star
  formation), these two types of galaxies could be essentially the same
  population of galaxies, sharing the same nature and origin. 

  Various possible mechanism are proposed to explain these phenomena.
  Poggianti et al. (1999) found passive spiral galaxies in their
  sample of distant clusters, and speculated that these findings show
  that the time scale of spectral changes of cluster galaxies is shorter
  than the time scale of morphological change of galaxies. They proposed
  ram-pressure stripping (Spitzer, Baade 1951; Gunn, Gott 1972;
  Farouki, Shapiro  1980; Kent 1981; Fujita 1998; Abadi et al. 1999;
  Fujita, Nagashima 1999; Quilis et al. 2000) as a possible physical
  mechanism responsible for these phenomena. Another possible cause is
  abrupt truncation of gas infall onto disks from the halo regions
  (Larson et al. 1980). Dynamical causes such as major galaxy merging 
  or harassment, which explain other properties of cluster
  galaxies very well (e.g., the Butcher--Oemler effect; Butcher, Oemler 1978, 
  1984), cannot explain these phenomena since such processes    
  disturb the spiral arms and do not end up with passive spirals.
  A pioneering work to simulate passive spiral galaxies by combining
  a numerical simulation and a phenomenological model was performed by Bekki
  et al. (2002). They demonstrated that halo gas stripping caused by a
  dynamical interaction between the halo gas and the hot ICM is a plausible
  mechanism.  
  Although these mechanisms are all plausible, no final conclusion
  about what mechanisms are responsible for these 
  phenomena has yet drawn.

  It is also interesting to investigate a possible link between passive
  spirals and statistical observational features of cluster galaxies. In
  cluster regions, it is known that fractions of  blue galaxies are
  larger at  higher redshifts. (the Butcher--Oemler effect; Butcher,
  Oemler 1978, 1984; Rakos, Schombert 1995; Couch et al. 1994,1998;
  Margoniner, de Carvalho 2000; Margoniner et al. 2001; Ellingson  et
  al. 2001; Kodama, Bower 2001; Goto et al. 2003a). Cluster galaxies are
  also known to change their morphology 
  during the cosmic timescale, e.g., spiral to S0 transition (Couch,
  Sharples 1987; Dressler 
  et al. 1997;  Couch et al. 1998; Fasano et al. 2000; 
 Diaferio et al. 2001)  or the
  morphological Butcher--Oemler effect (Goto et al. 2003a). If passive spiral
  galaxies are cluster-originated, they might fit well in both spectral
  and morphological evolution of cluster galaxies, as galaxies in transition
  between blue and red, or spiral and S0s.

 Since the Sloan Digital Sky Survey (SDSS; York et al. 2000) observes
 the spectra of one million galaxies in one quarter of the sky,
 It provides
 us with the opportunity to study this interesting population of galaxies
 in all environments, from cluster core regions to general field
 regions. Several of the largest cluster catalog are compiled using the
  SDSS data (Annis et al. 1999; Kim et al. 2002; Goto et al. 2002a;
 Miller et al. in preparation).
 In addition, a wide spectral coverage of 3800--9000 \AA\ allows us
 to study both [O{\sc ii}] and H$\alpha$ emission lines at the same time,
 which can reduce 
 possible biases from dust extinction and stellar absorption on the
 emission lines. 
 In this paper, we concentrate on revealing the environment of passive
 spiral galaxies. In section 2, we explain the data used in the
 study. In section 3, we carefully define passive spiral galaxies. In
 section 4, we present the environment of passive spiral galaxies. In
 section 5, we discuss the possible caveats and interpretation of the
 results. In section 6, we summarize our findings. The cosmological
 parameters adopted throughout this paper are $H_0$=75 km
 s$^{-1}$ Mpc$^{-1}$, and ($\Omega_{\rm
 m}$,$\Omega_{\Lambda}$,$\Omega_{\rm k}$)=(0.3,0.7,0.0).

\section{Data}\label{data}

  In this section, we outline the data used in this paper. 
  The galaxy catalog is taken from the Sloan Digital Sky Survey (SDSS;
  for more details,
  see Fukugita et al. 1996; Gunn et al. 1998;  Lupton 
  et al. 1999, 2001; York et al. 2000; Eisenstein et al. 2001; Hogg
  et al. 2001; Richards et al. 2002; Stoughton et al. 2002; Strauss et
  al. 2002; Smith et al. 2002; Blanton et al. 2003a; Pier et al. 2003).
  The SDSS imaging survey observes one quarter of the sky to depths of
  22.3, 23.3, 23.1, 22.3, and 20.8 in the $u,g,r,i,$ and $z$ filters,
  respectively (see Fukugita et al. 1996 for the SDSS filter system,
  Hogg et al. 2001 
 and Smith et al. 2002 for its calibration). 
 Since the SDSS photometric system is not yet finalized, we refer to the
 SDSS photometry presented here as $u^*,g^*,r^*,i^*,$ and $z^*$. We
  correct the data for galactic extinction determined from maps
  given by Schlegel, Finkbeiner, and Davis (1998).  
 We include galaxies to $r^*$=17.7 (Petrosian magnitude), which is the
 target selection limit of the main galaxy sample of the SDSS spectroscopic survey.
 The spectra are obtained using two fiber-fed spectrographs (each with 320
 fibers) with each fiber subtending 3 arcsec on the sky. 
 (We investigate aperture bias due to the limited size of the SDSS fiber
 spectrograph in the appendix). 
 The wavelength
 coverage of the spectrographs is 3800 \AA\ to 9200 \AA, with a spectral
 resolution of 1800. These spectra are then analyzed via the SDSS
 SPECTRO1D data-processing 
 pipeline to obtain various quantities for each spectrum such as
 redshift, spectral classification, and various line parameters. (see
  Stoughton et al. 2002; Frieman et al. in preparation, for further 
 details).  
  The SDSS has taken 189763 galaxy spectra as of the date of writing. Among
  them we restrict our sample to galaxies with S/N in $g$ band
  greater than 5 and with a redshift confidence of $\geq0.7$.
 Since we use a concentration parameter in selecting passive
 spiral galaxies, we also remove galaxies with a PSF size in $r$ band
 greater than 
 2.''0 to avoid poor seeing mimicking less concentrated galaxies.  
  Then we make a volume-limited sample by restricting our sample to
 0.05$<z<$0.1 and  $Mr^*<-20.5$. This magnitude limit corresponds to
 $Mr^*$+0.3 mag (Blanton 
  et al. 2001).  
  The lower redshift cut is made to avoid strong aperture
  effects (see appendix for detailed investigation in aperture
  effects). When calculating absolute magnitudes,  
  we use a  $k$-correction code provided by Blanton et al. (2003b; v1\_11). 
  In this volume-limited sample, there are 25813 galaxies. 

\section{Selection of Passive Spiral Galaxies}

\subsection{Line Measurements}
 
 We measured [O{\sc ii}], H$\alpha$ and H$\delta$ equivalent widths (EWs) by
 the flux-summing method 
 as described in Goto et al. (2003b). We briefly summarize the method
 here.  To estimate the continuum, we fit a line using a wavelength range
 around each line as listed in table
 \ref{tab:wavelength}. The continuum values were weighted according to the
 inverse square of the errors during the fitting procedure. We then sum
 the flux in the wavelength range listed in the same table to obtain the
 equivalent width of the lines. For  the H$\delta$ line, we used two
 different wavelength ranges: wider one for a strong line and narrower one
 for a weak line (details are described in Goto et al. 2003b). 
 Note that for the H$\alpha$ line, we did not
 deblend the adjacent [N{\scshape ii}] lines. As a result, our H$\alpha$ equivalent
 width had contamination from the [N{\scshape ii}] lines. However, the contamination was
 less than 5\% from [N{\scshape ii}](6648 \AA) and less than 30\% from
 [N{\scshape ii}](6583 \AA).  These measurements show good
 agreement with measurements via Gaussian fitting (Goto et al. 2003b). 
 We, however, stress the importance in using the flux-summing method
 instead of a Gaussian-fitting method, which is  often used in other
 work (e.g., G{\' o}mez et al. 2003). Although a Gaussian fitting method works very well for
 strong lines on high signal-to-noise spectra, the fit often fails on low signal-to-noise spectra, 
 especially when the line is weak. Therefore, a Gaussian-fitting method
 is not suitable to study passive spiral galaxies, which intrinsically
 have little emission lines. On the other hand, the flux-summing method
 is less affected by the noise of the spectra, and thus, is suitable for
 this study. 

 We quantified errors of these measurement using spectra observed
 twice in the SDSS. The procedure was exactly the same as described in
 Goto et al. (2003b). First, the difference of equivalent width was
 plotted against S/N of the spectra . We then fit a 3rd polynomial to the 1
 $\sigma$ of the distribution. The polynomial was later used to assign
 errors to every spectra according to its S/N. The exact formula are
 given in Goto et al. (2003b). Typical errors of high signal-to-noise
 spectra are  1.3, 1.0 and 0.4 \AA\
 for [O{\sc ii}], H$\alpha$ and H$\delta$ EWs, respectively (See Figures 9--11 of
 Goto et al. 2003b).

\subsection{Selection Criteria}\label{selection}

\begin{figure}
\centering{\includegraphics[scale=0.4]{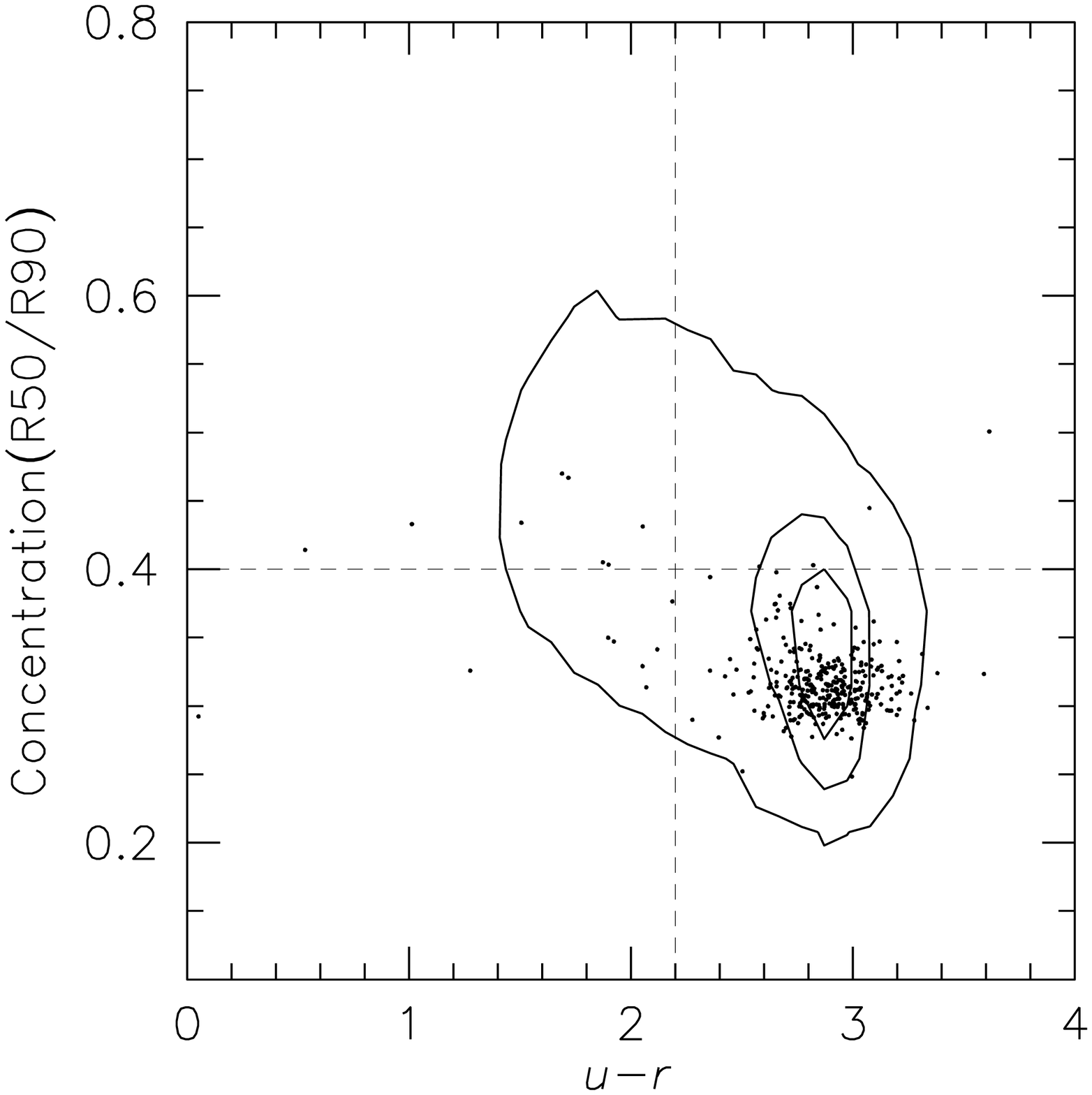}
\includegraphics[scale=0.4]{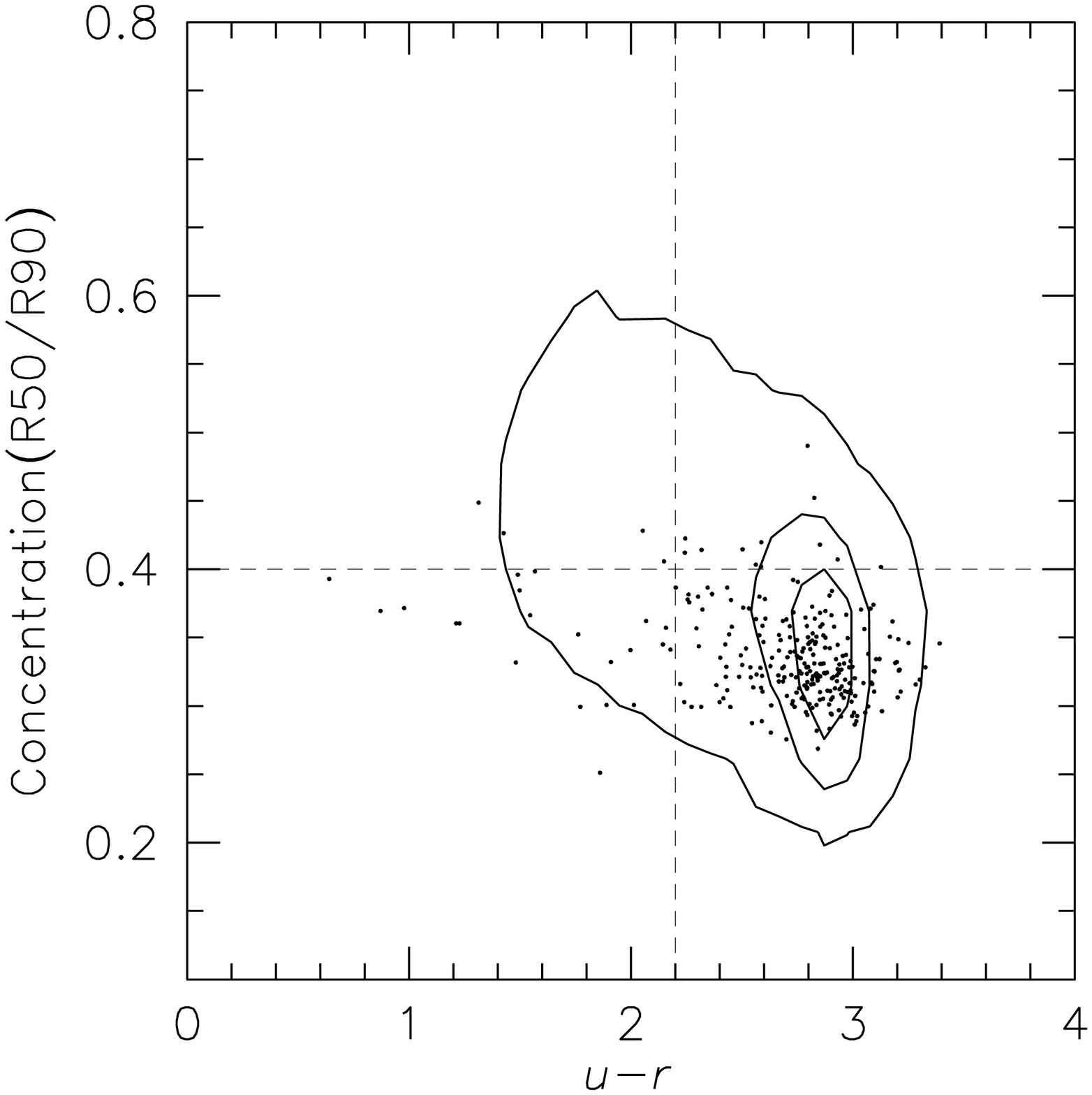}}
\centering{\includegraphics[scale=0.4]{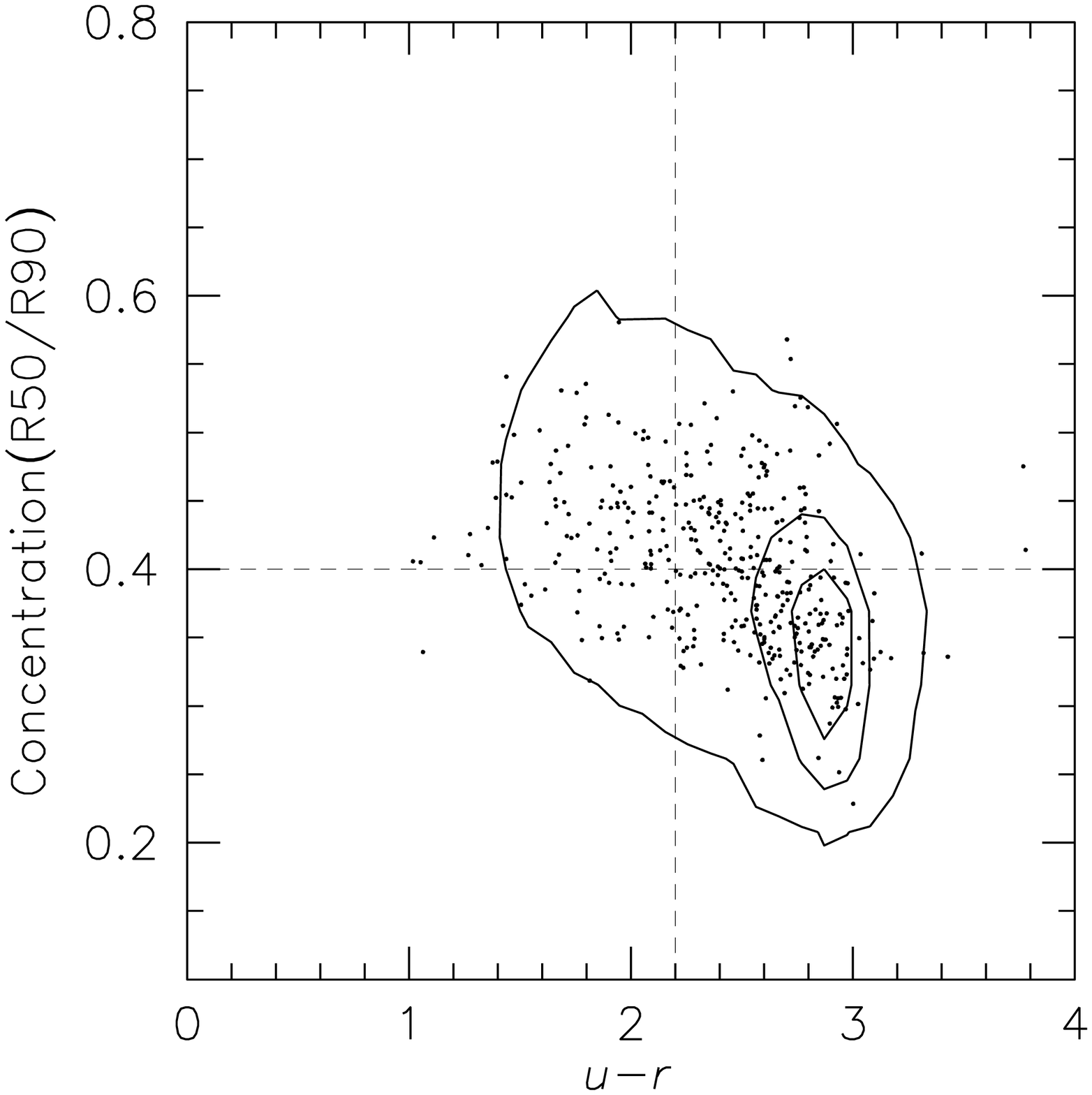}
\includegraphics[scale=0.4]{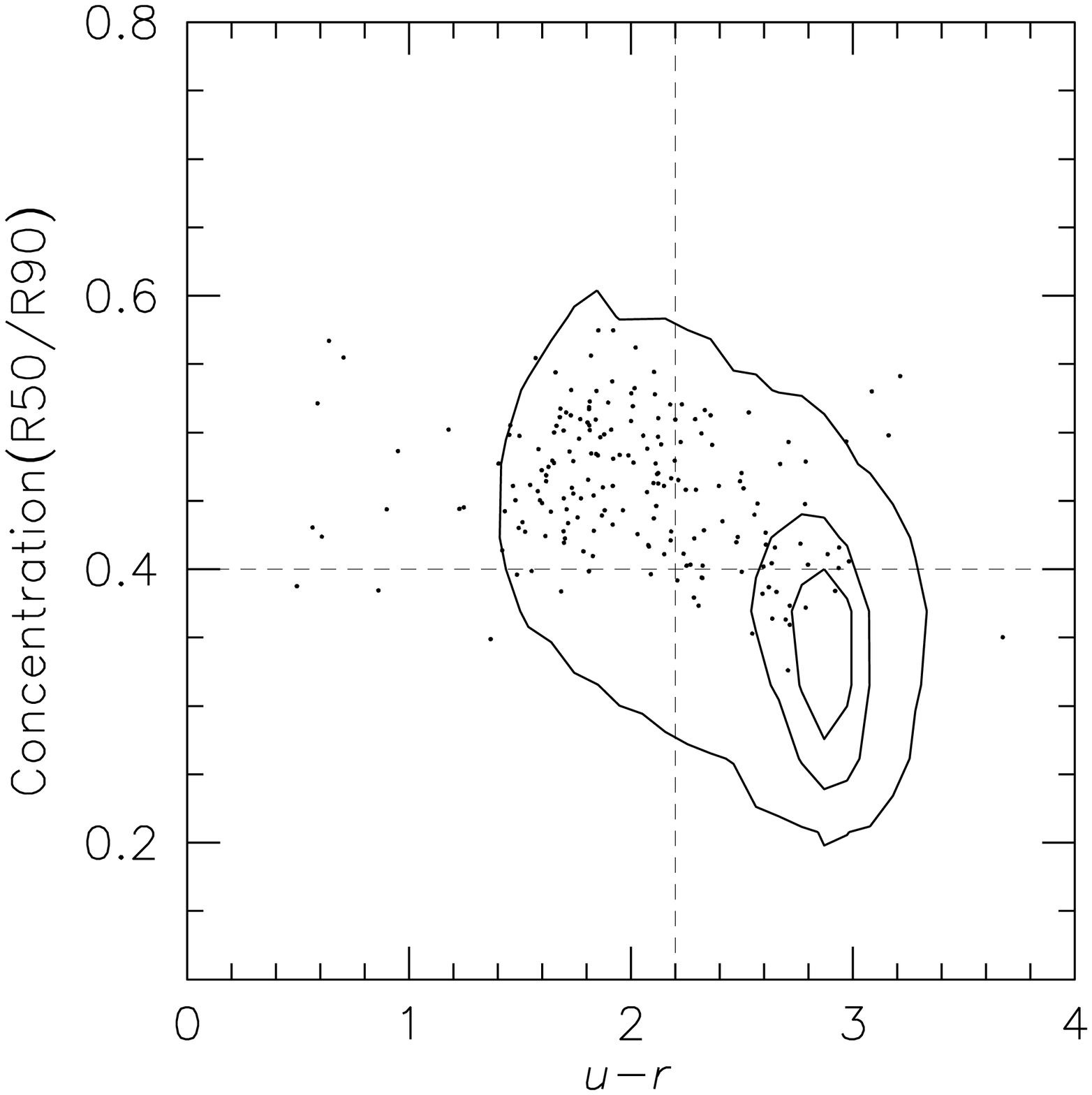}}
\caption{
 $C_{\rm in}$ is plotted against $u-r$. The contours show the distribution of all
 galaxies in the volume-limited sample (0.05$<z<$0.1 and
  $Mr^*<-20.5$). A good correlation between the 
 two parameters is seen. The points in each
 panel show the distribution of each morphological type of galaxies
 classified by eye (Shimasaku et al. 2001; Nakamura et
 al. 2003); Ellipticals are in the upper left panel. S0, Sa and Sc are
 in the upper right, lower left and lower right panels, respectively.   
}\label{fig:concent_each_type}
\end{figure}

 We selected passive spiral galaxies using the following criteria. 
 Galaxies with an inverse of concentration parameter, $C_{\rm in}>$0.5. 
 The inverse concentration parameter ($C_{\rm in}$) is defined as the
   ratio of the Petrosian 50\% light radius to the Petrosian 90\% light radius
 in $r$ band 
 (radius which contains 50\% and 90\% of Petrosian flux, respectively).
 Shimasaku et  al. (2001) and Strateva et al. (2001) studied the
 completeness and 
 contamination of this parameter in detail. See Goto et al. (2003a) and
  G{\' o}mez et al. (2003) for more usage of this parameter. The border line
 between spiral galaxies and elliptical galaxies are around
 $C_{\rm in}$=0.33. Therefore $C_{\rm in}>$0.5 can select very
 less-concentrated  spiral galaxies. 
 In different work, 549 galaxies in our volume-limited sample
 were manually classified by Shimasaku et al. (2001) and Nakamura et al.
 (2003). In figure \ref{fig:concent_each_type}, we overplot
 eye-classified galaxies on a $C_{\rm in}$ vs. $u-r$ plane. The contours
 show the 
 distribution of all galaxies in our volume-limited sample. Note that
 our volume-limited sample contains a high fraction of concentrated
 galaxies as shown by contours
 due to its bright absolute magnitude limit (Yagi et al. 2002; Goto et
 al. 2002b).
 In the top
 left panel, 
 eye-classified ellipticals are overplotted. In the top right, bottom
 left, bottom right panels, eye-classified S0s, Sa-Sb, Sc or later are
 overplotted, respectively. As shown in Strateva et al. (2001),
 $u-r=2.2$ also separates early and late-type galaxies well.   
  To exclude contamination from S0 galaxies as much as possible, we adopted stricter
 criteria ($C_{\rm in}>$0.5) than the $C_{\rm in}>$0.4 criteria often used in
 other work (e.g.,  G{\' o}mez et al. 2003; Goto et al. 2002b, 2003a). 
 As these panels show, few of elliptical or
 S0 galaxies have $C_{\rm in}>0.5$.
 Therefore, we in fact were able to select
 spiral galaxies using $C_{\rm in}$ parameter, without significant
 contamination from the E/S0 population. We caution readers that the
 selection of less-concentrated galaxies has a known bias against
 edge-on galaxies, in the sense that edge-on disc galaxies are excluded
 from our sample. A detailed investigation and correction of this
 bias will be presented in Yamauchi et al. (in preparation). However, we accept
 this bias in our sample selection since (i) the bias is independent of
 the local
 galaxy environment; (ii) edge-on galaxies might be affected by a larger
 amount of dust extinction, and thus could cast some doubts on the truly
 passive nature of our sample galaxies.

\begin{figure}
\begin{center}
\includegraphics[scale=0.7]{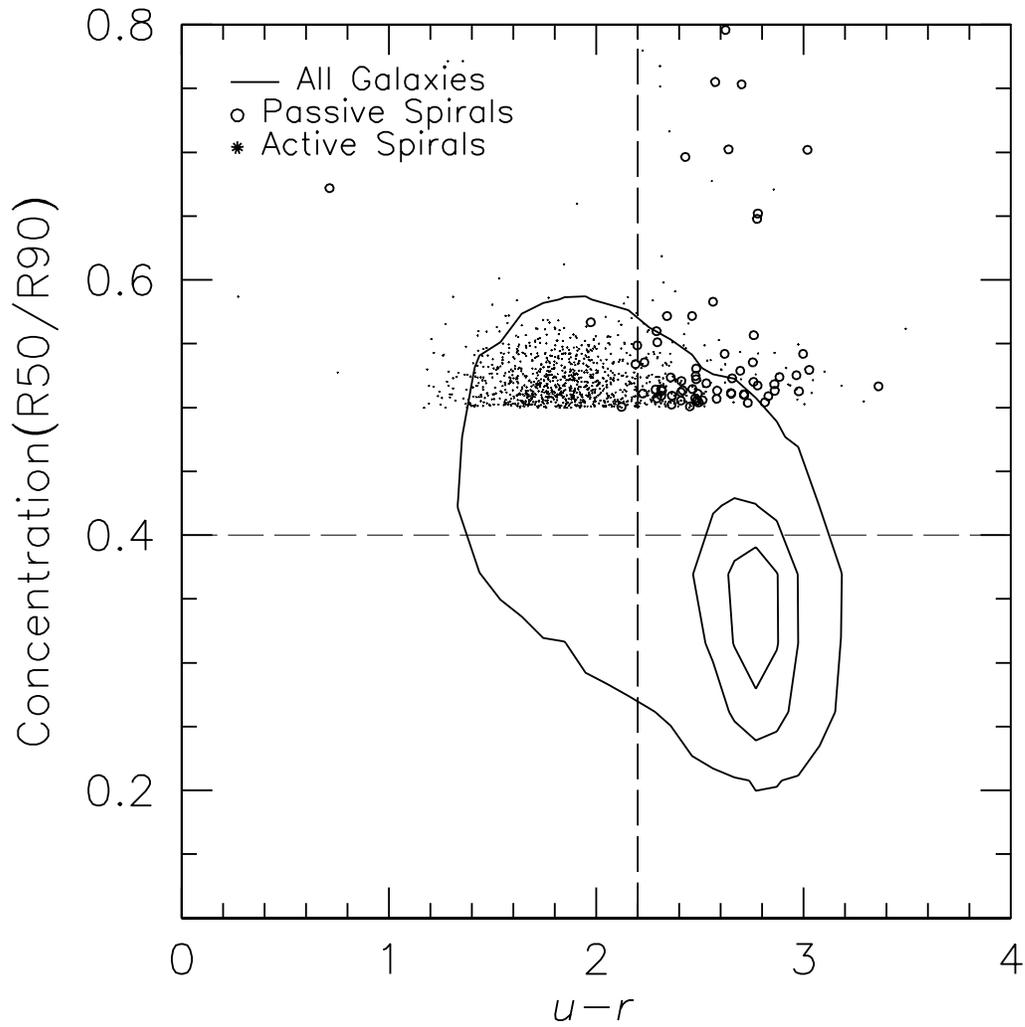}
\end{center}
\caption{
\label{fig:ps_concent}
 Distribution of passive spiral galaxies in $C_{\rm in}$ vs. $u-r$
 plane. The contours show the distribution of all galaxies in our
 volume-limited sample. The open circle and filled dots represent passive
 and  
 active spiral galaxies, respectively.}
\end{figure}

   From the spectral features, we select galaxies using the following
   criteria:\\
  \begin{equation} 
  [{\rm OII}]\ EW - 1\sigma_{error}< 0, 
 \end{equation}
  \begin{equation} 
  H\alpha\ EW - 1\sigma_{error}< 0,   
 \end{equation}
  where the emission lines have positive signs. 
 In other words, we selected galaxies with [O{\sc ii}] and H$\alpha$
 less than 1 $\sigma$ detection (in emission).
 A galaxy that satisfies both the concentration and spectral
 criteria was regarded as being a passive spiral galaxy in this work.
  An alternative way is to use the star-formation rate (SFR). However,, 
 since passive spirals intrinsically have little emissions in [O{\sc ii}] and H$\alpha$, it
 is erroneous to calculate SFR based on [O{\sc ii}] or H$\alpha$.  Therefore it
 is clearer to select a sample based on the detection or non-detection
 of emission lines.
  Figure \ref{fig:ps_concent} 
 shows the distribution of passive spiral galaxies in the $C_{\rm in}$
 vs. $u-r$ plane.  
 
\begin{figure}
\centering{
\includegraphics[scale=1.]{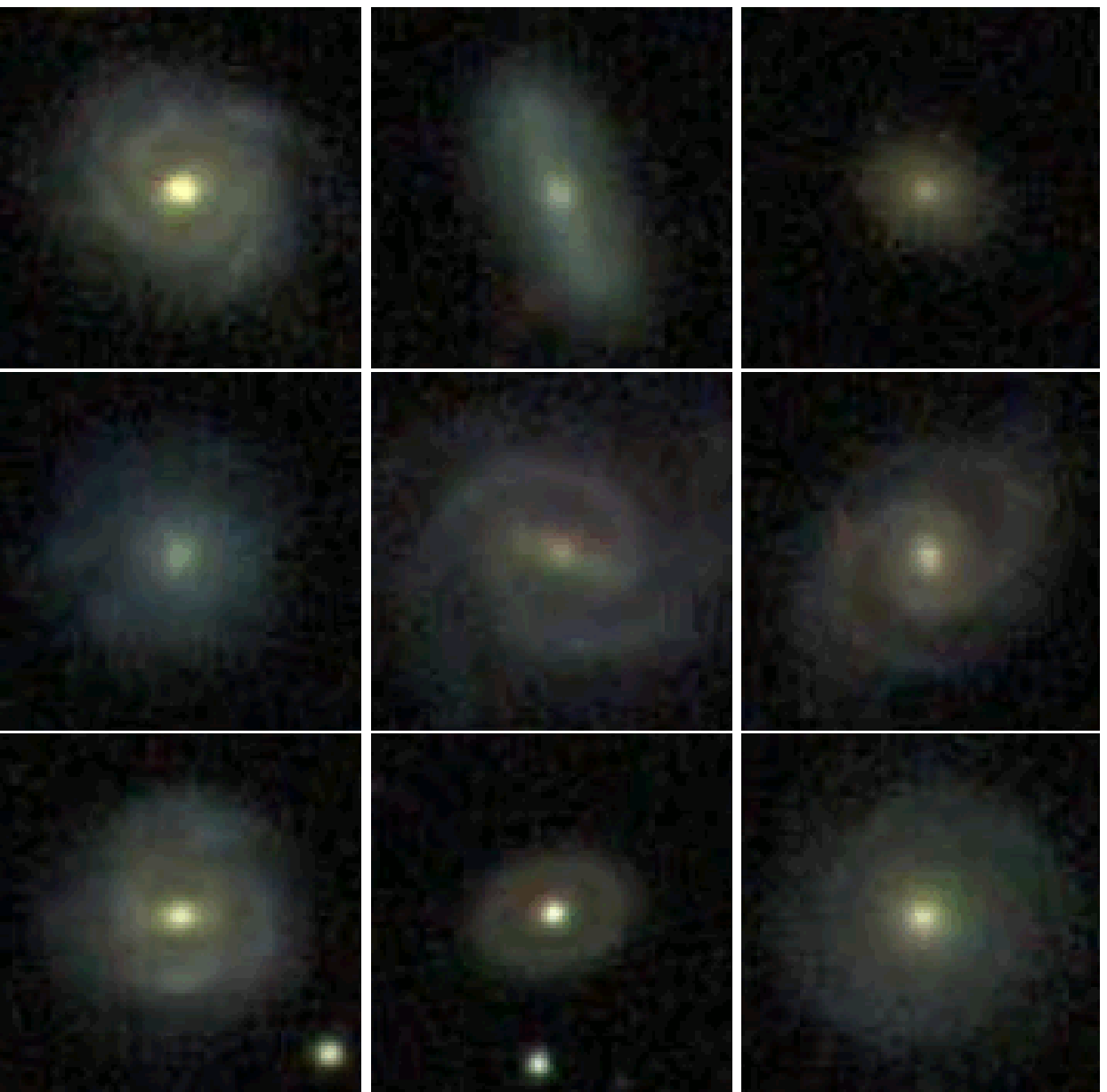}
}
\caption{
\label{fig:image}
 Example images of passive spiral galaxies. Each image is a composite of
 SDSS $g,r,$ and $i$ bands, showing a
 30''$\times$30'' area of the sky with its north up.
 Discs and spiral arm structures can be recognized.
}\end{figure}

\begin{figure}
\centering{
\includegraphics[scale=0.25]{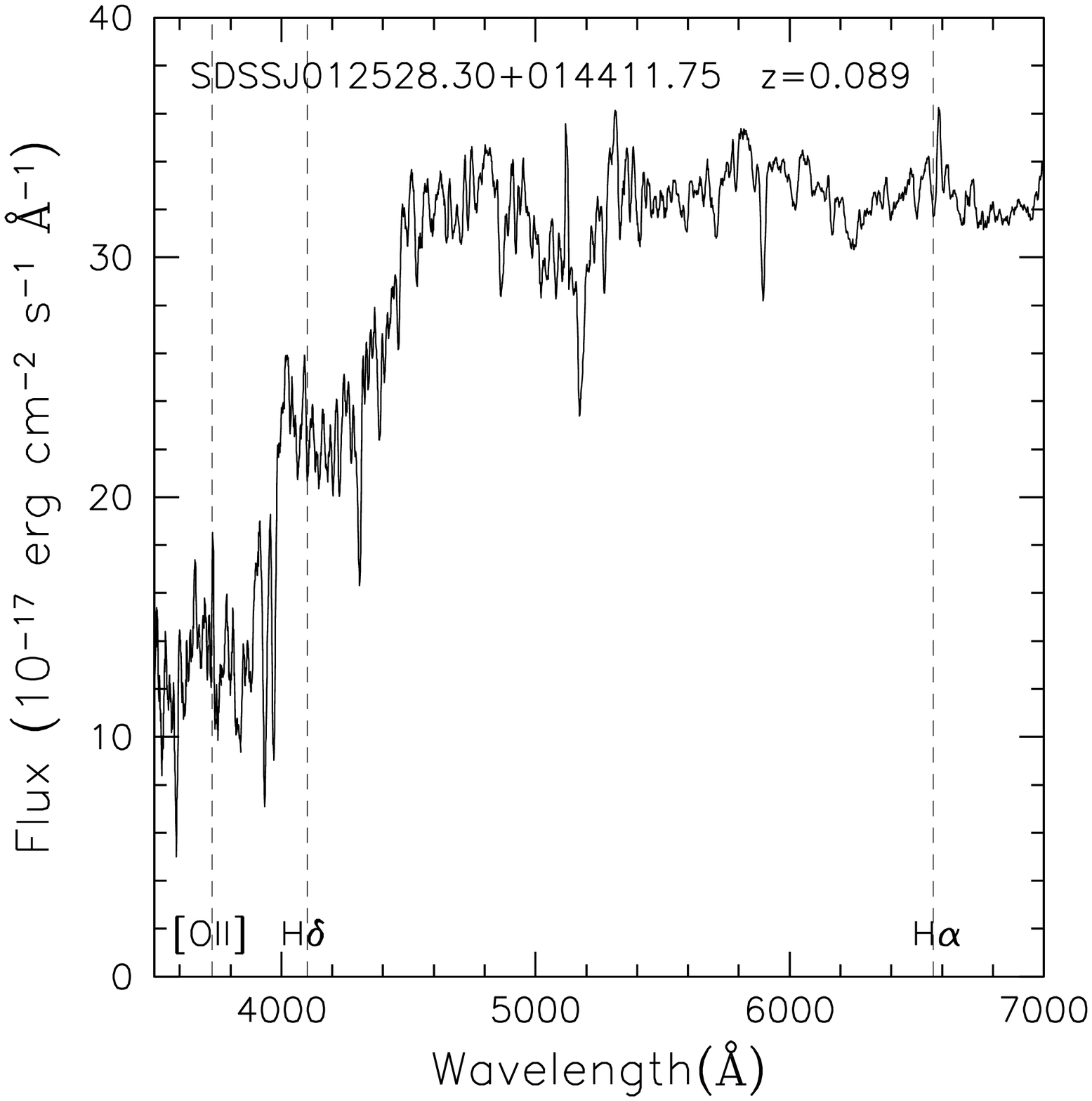} %
\includegraphics[scale=0.25]{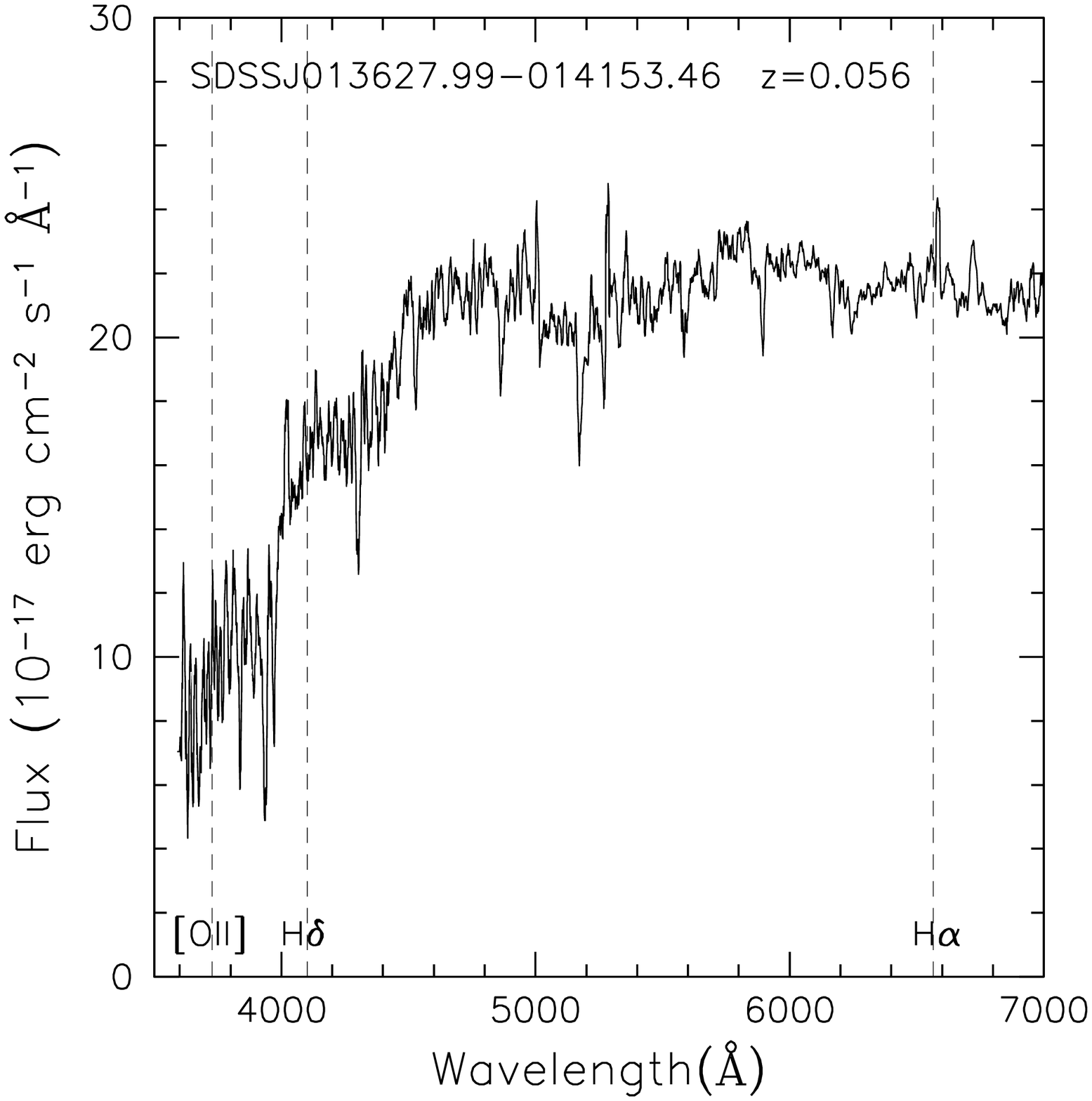}
\includegraphics[scale=0.25]{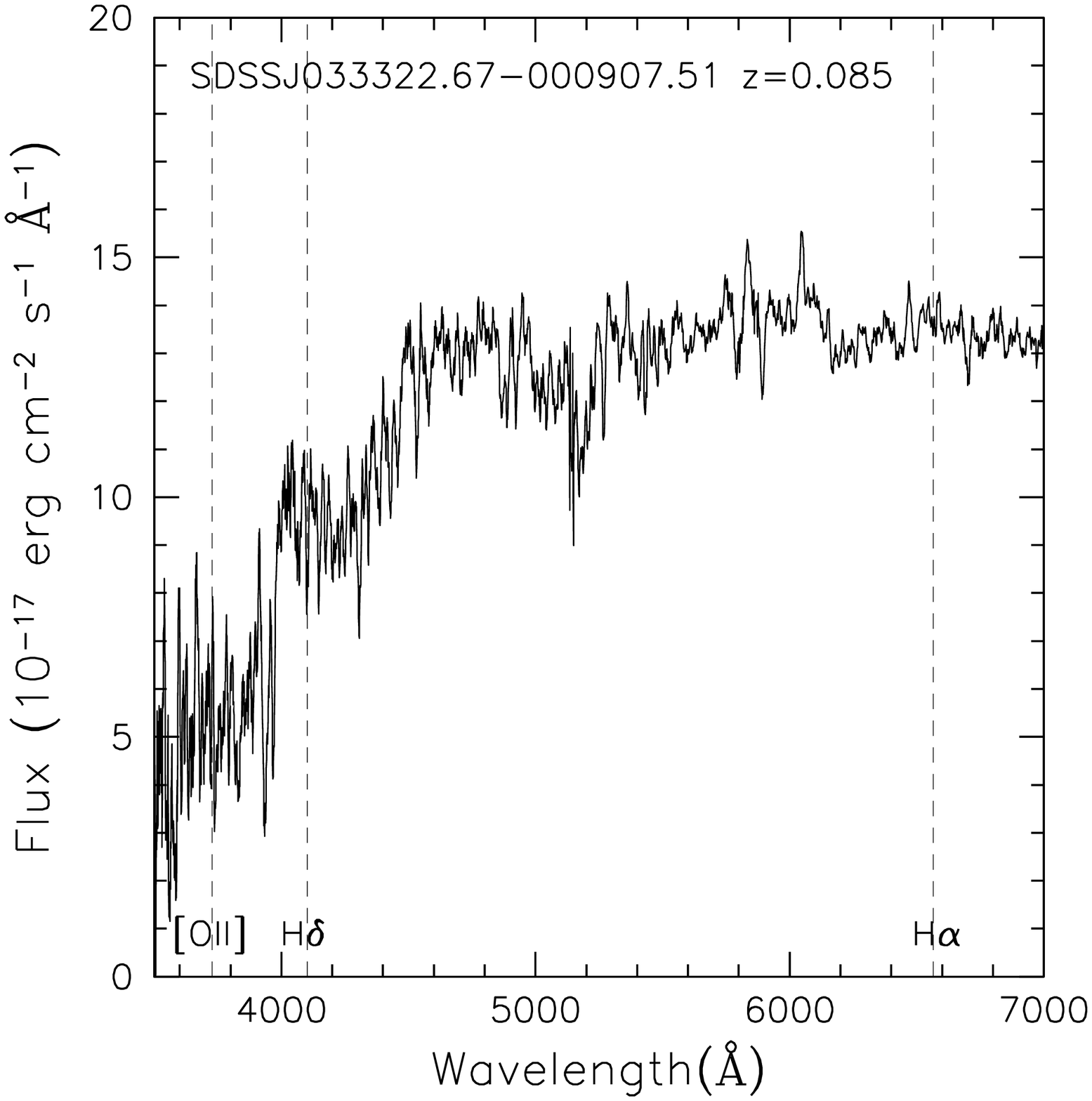}
}
\centering{
\includegraphics[scale=0.25]{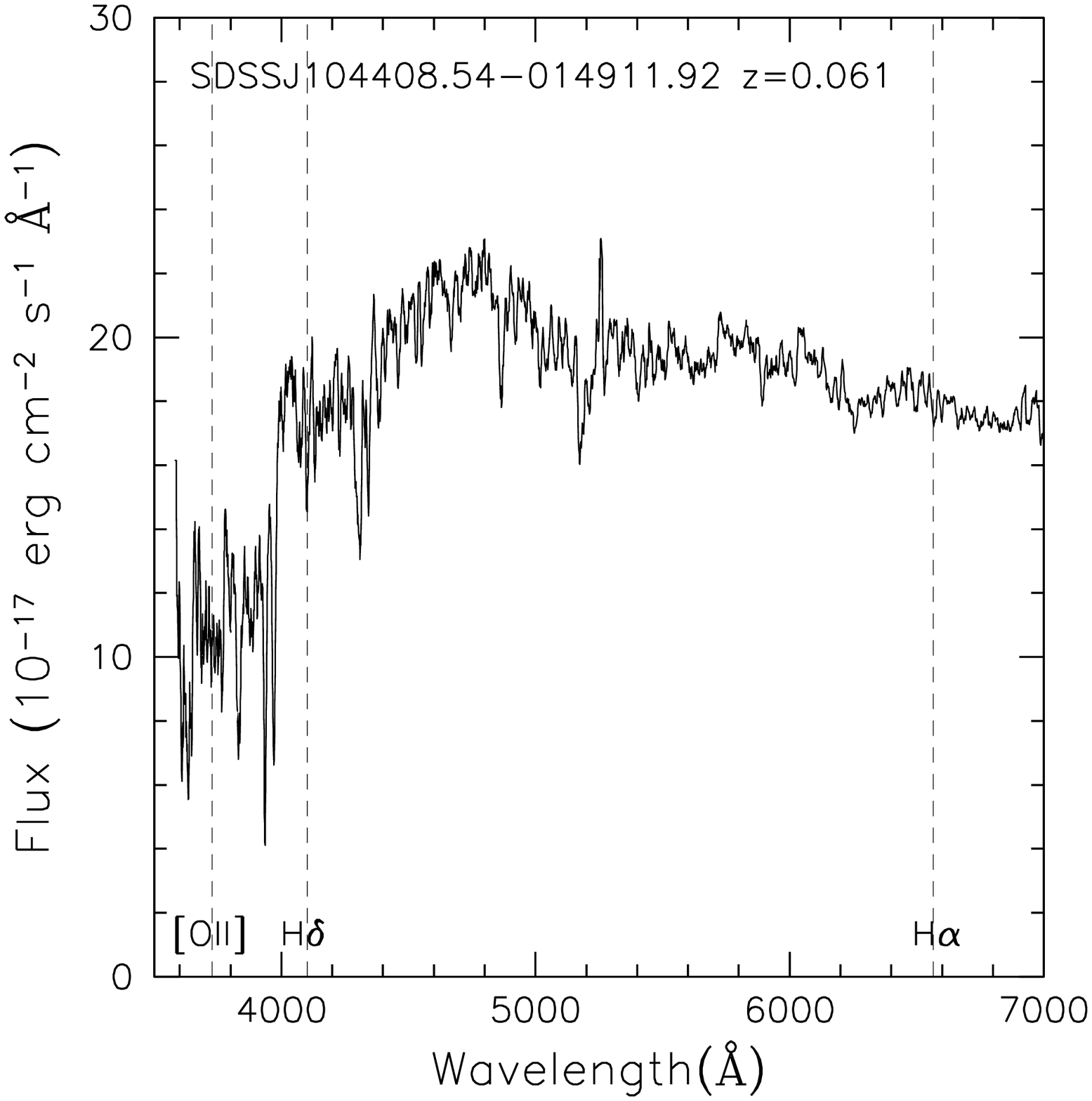}
\includegraphics[scale=0.25]{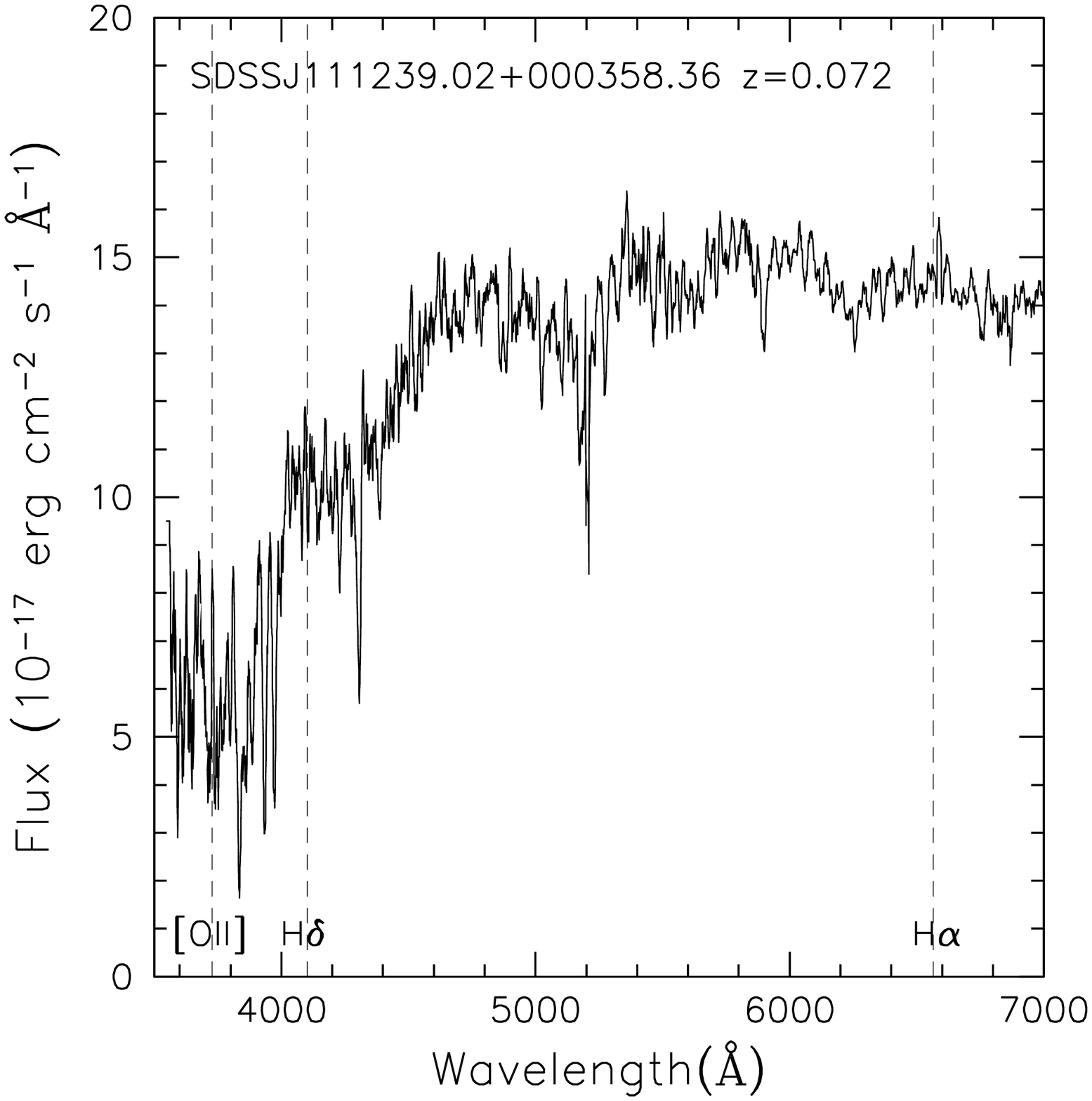}
\includegraphics[scale=0.25]{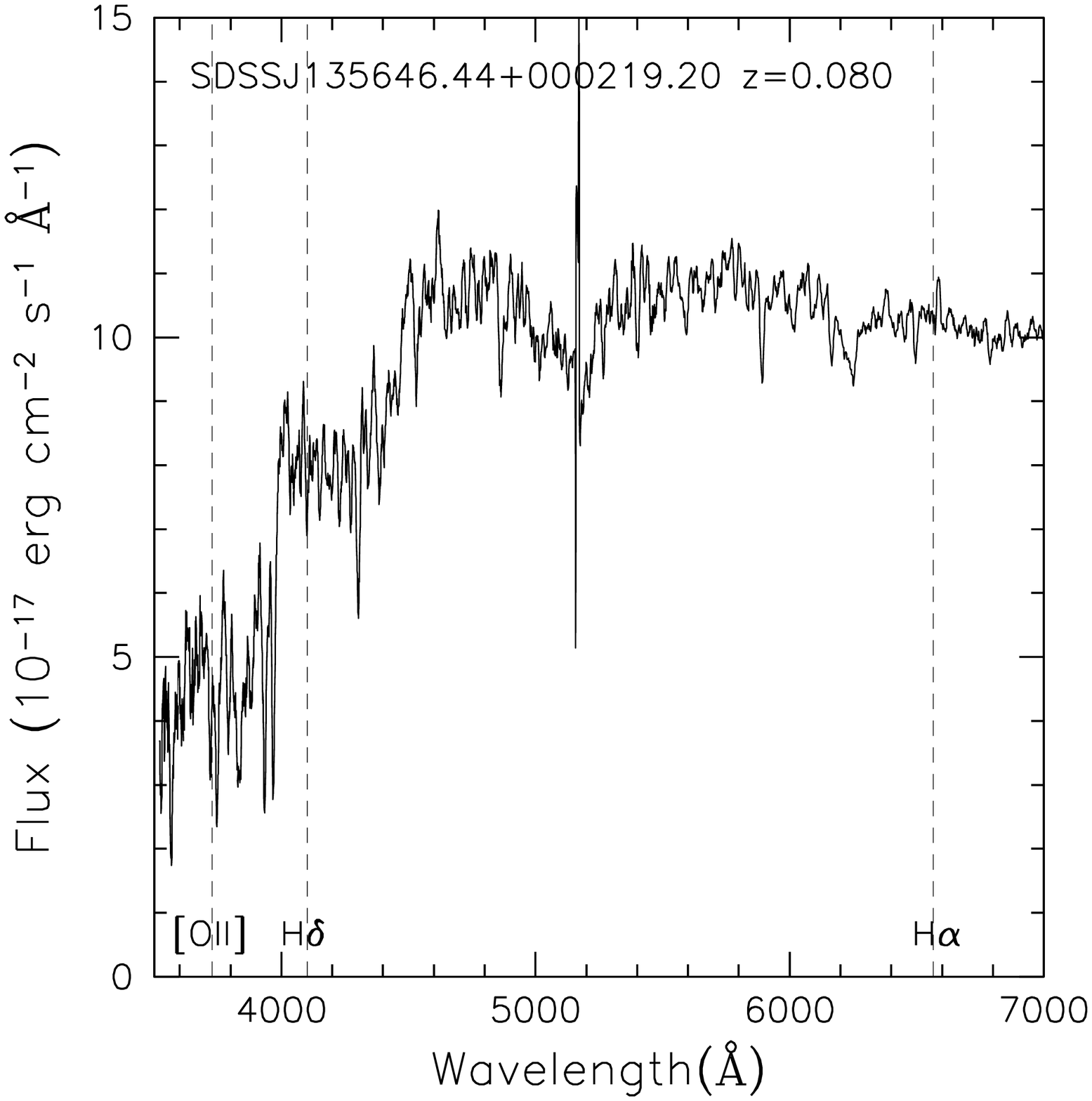}
}
\centering{
\includegraphics[scale=0.25]{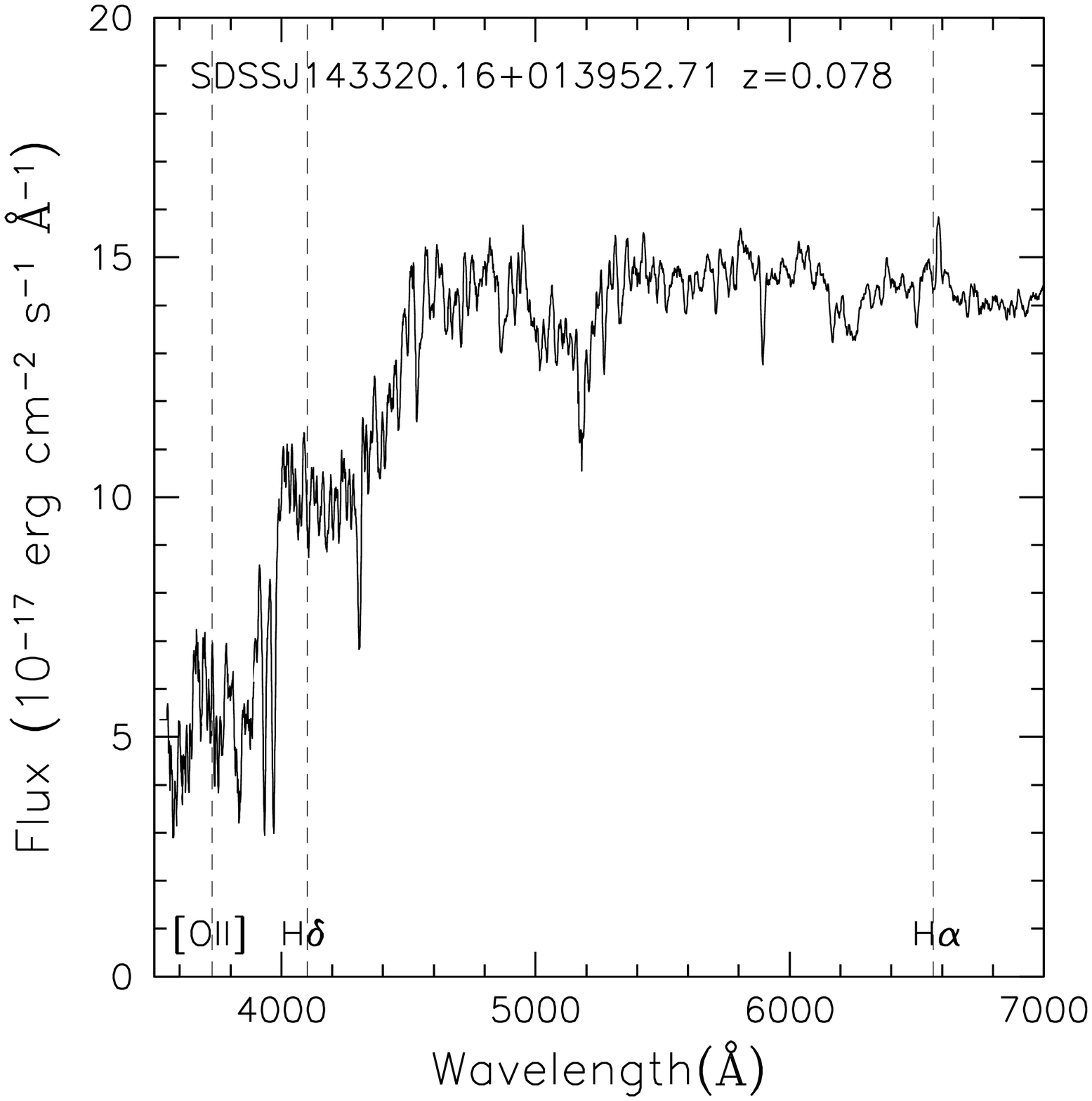}
\includegraphics[scale=0.25]{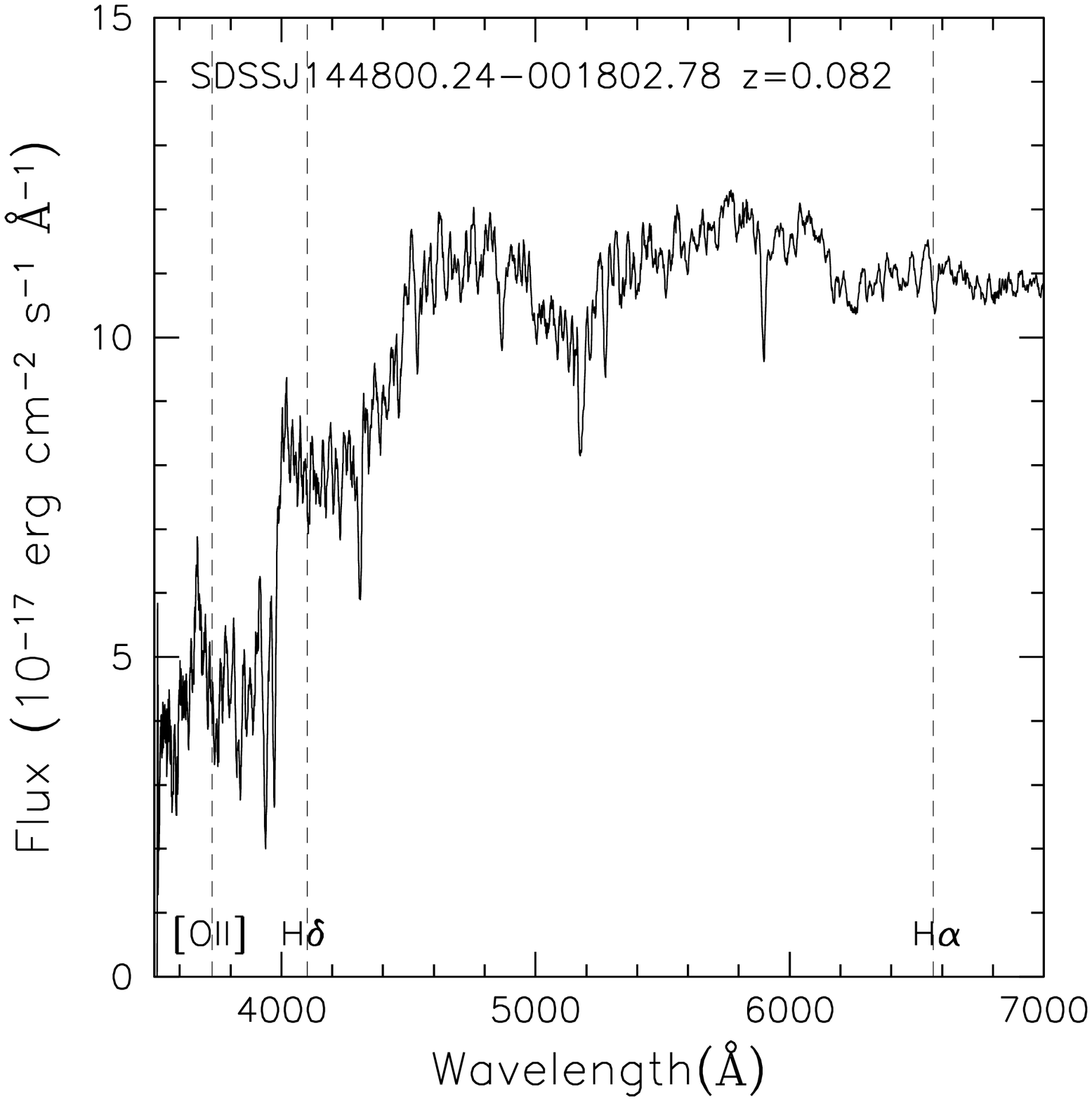}
\includegraphics[scale=0.25]{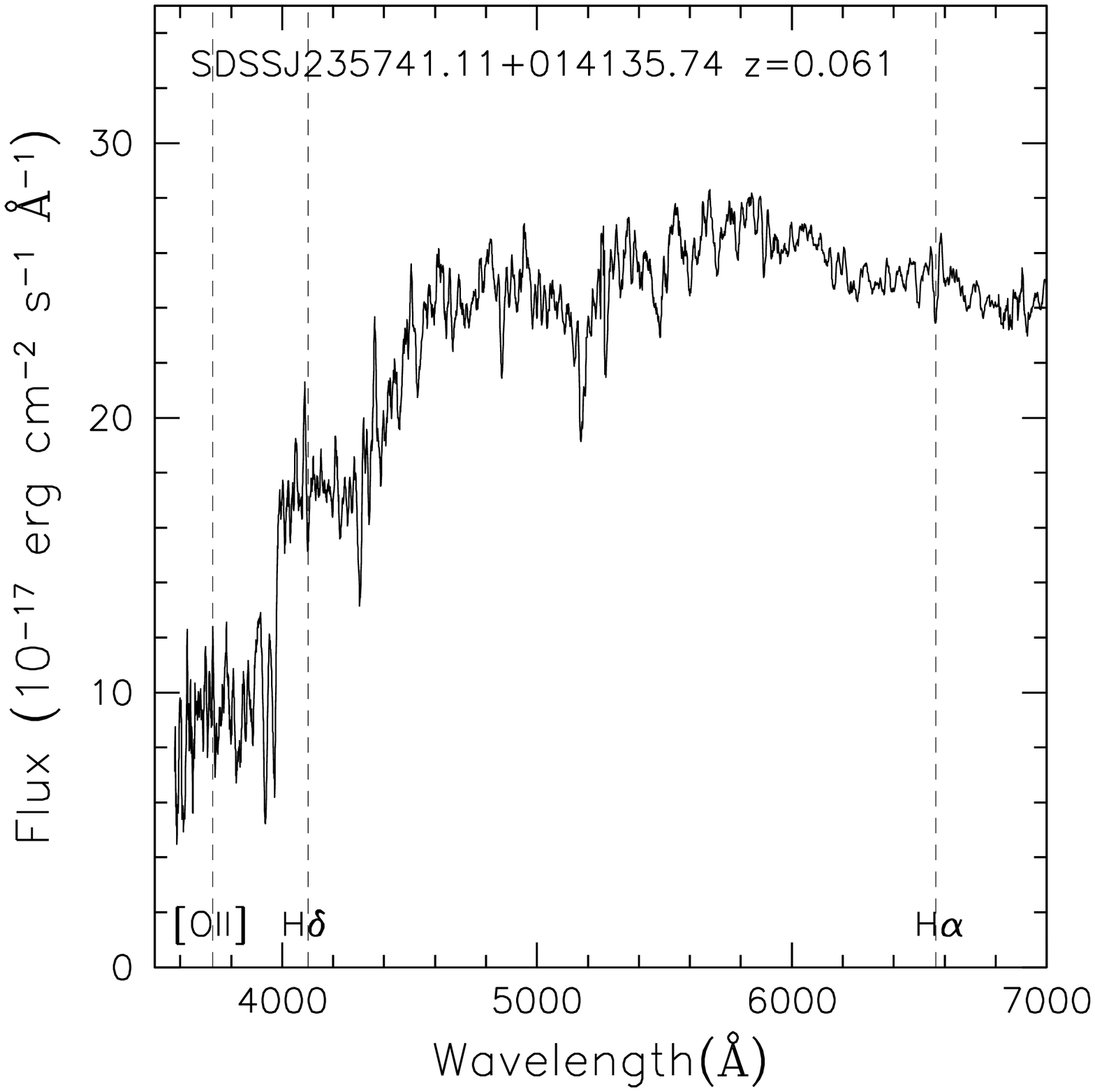}
}
\caption{
\label{fig:spectra}
 Example rest-frame spectra of passive spiral galaxies. The spectra are
 shifted to rest-frame and smoothed using a 10 \AA\ box.
 Each panel corresponds to
 that in figure \ref{fig:image}. 
}\end{figure}

\begin{figure}
\centering{
\includegraphics[scale=1.]{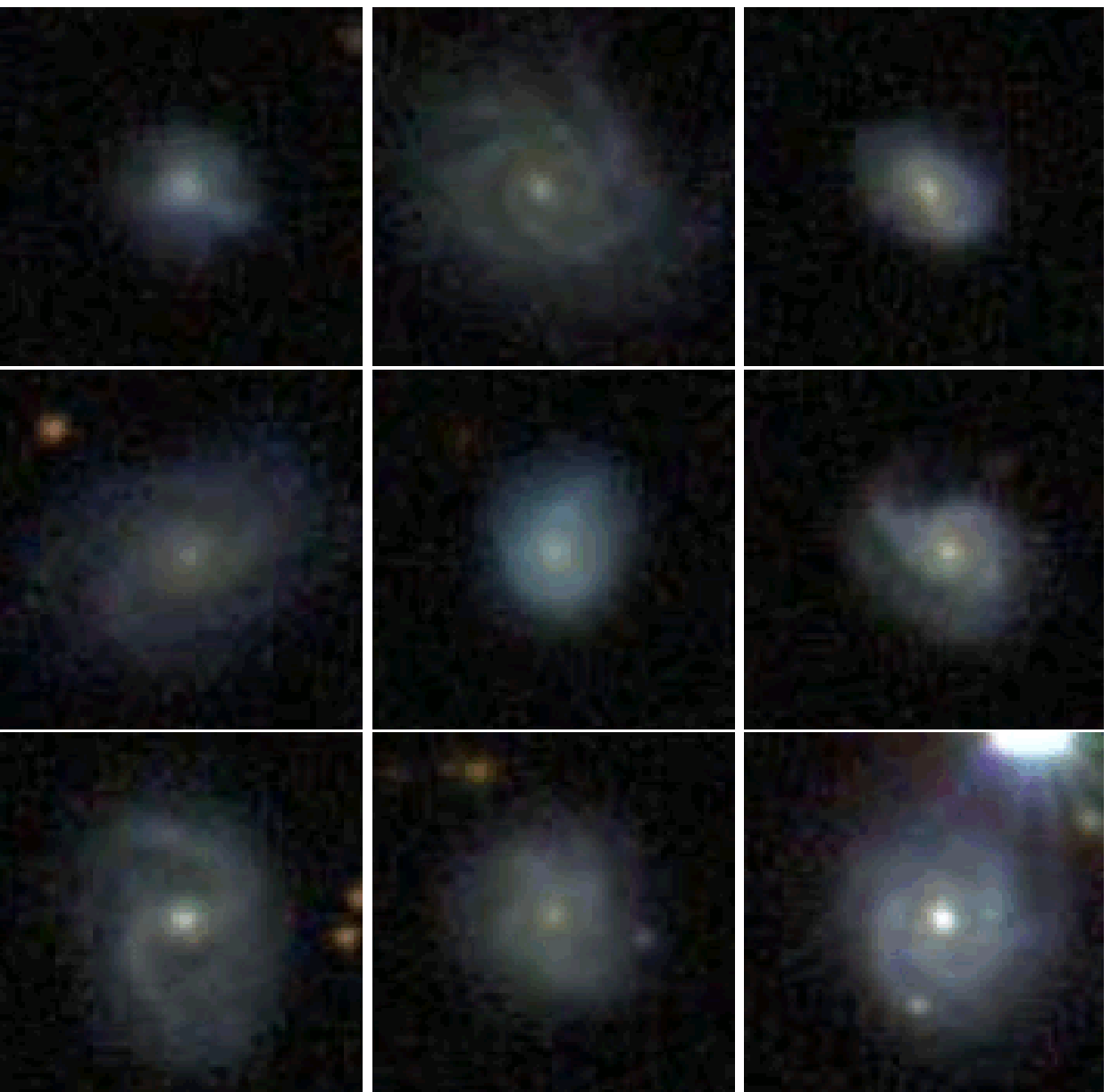}
}
\caption{
\label{fig:as_image}
 Example images of active spiral galaxies. Each image is a composite of
 SDSS $g,r,$ and $i$ bands, showing
 a 30''$\times$30'' area of the sky with its north up.
 Discs and spiral arm structures can be recognized.
}\end{figure}

\begin{figure}
\centering{
\includegraphics[scale=0.25]{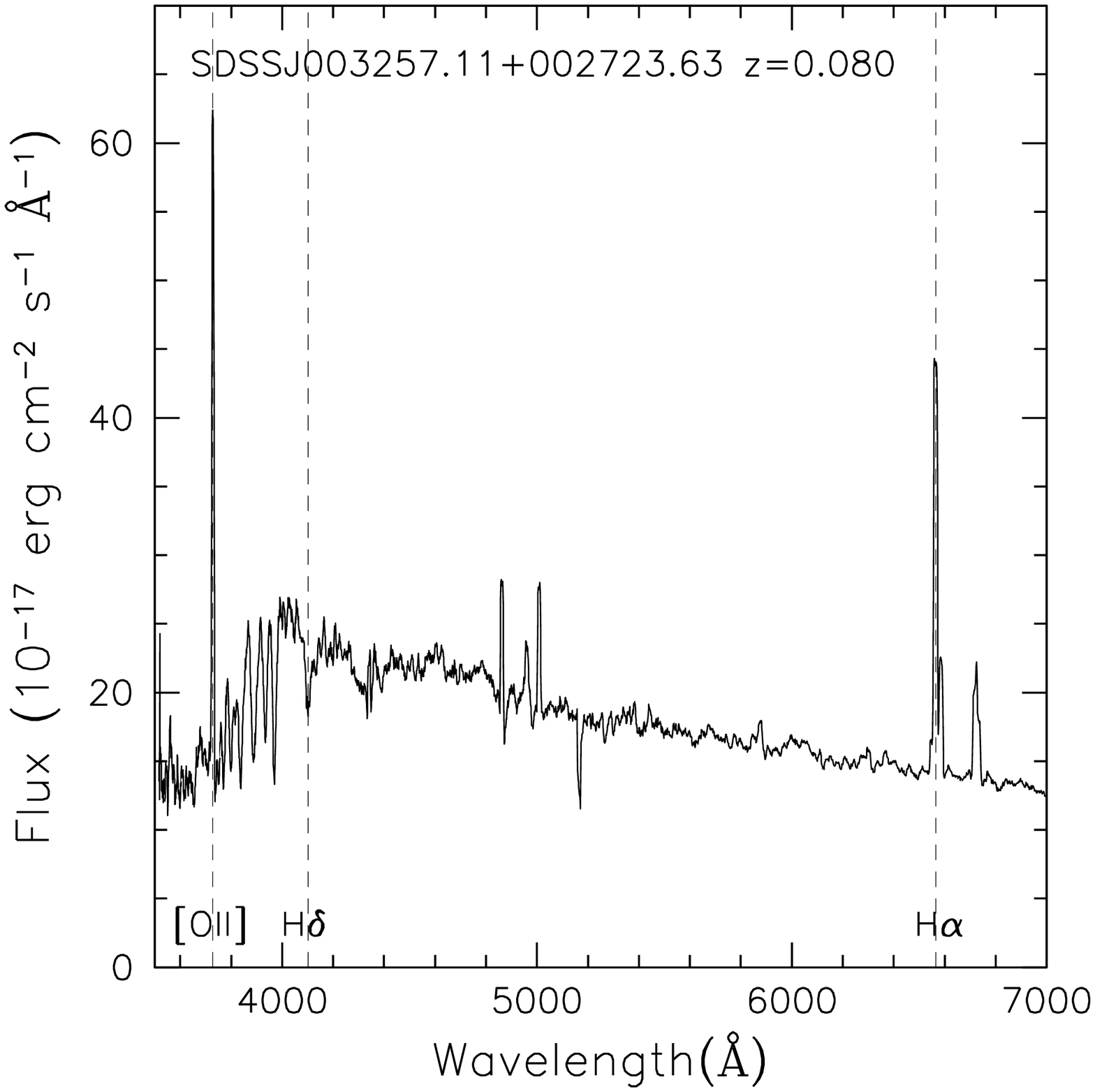} %
\includegraphics[scale=0.25]{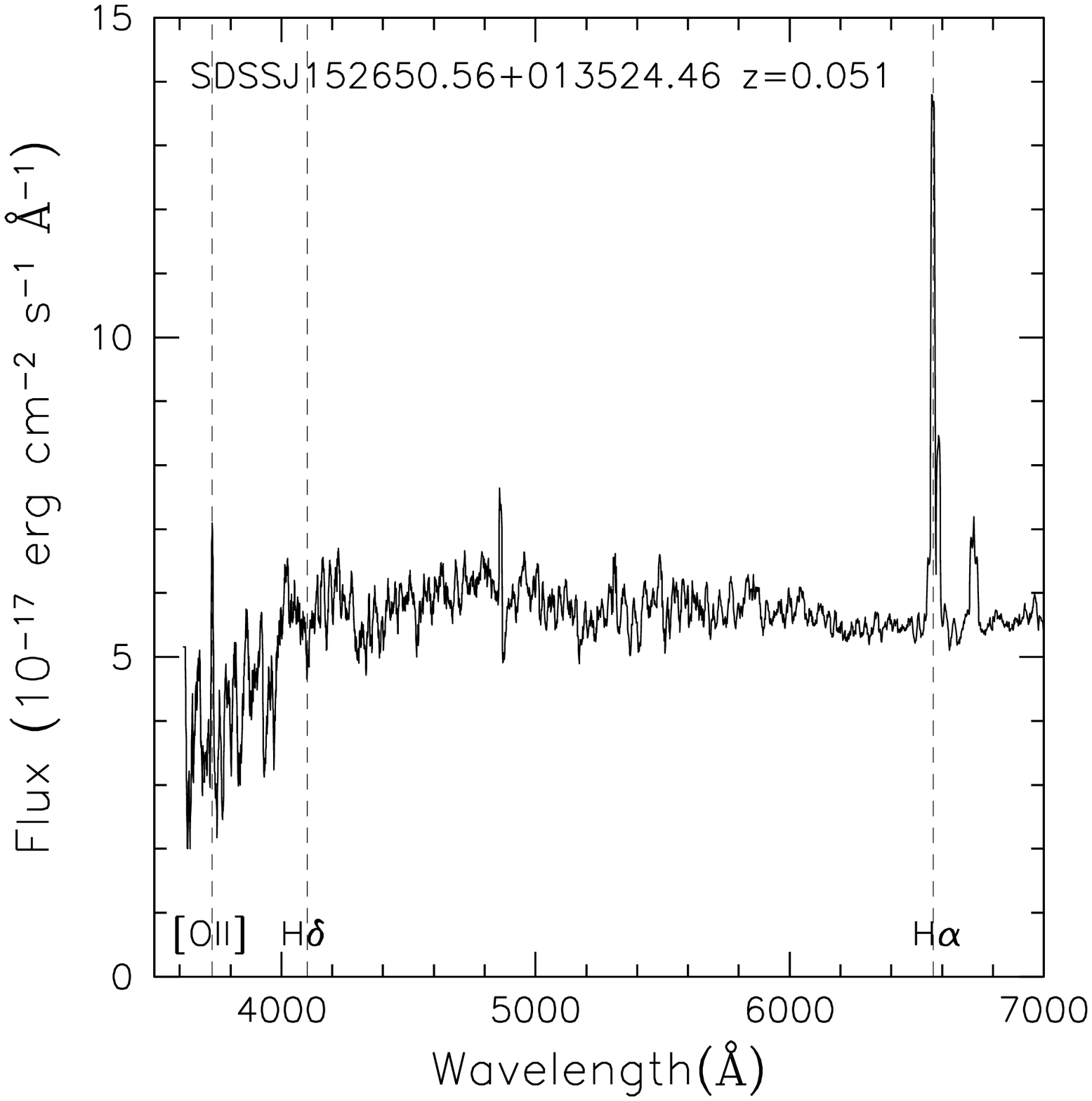}
\includegraphics[scale=0.25]{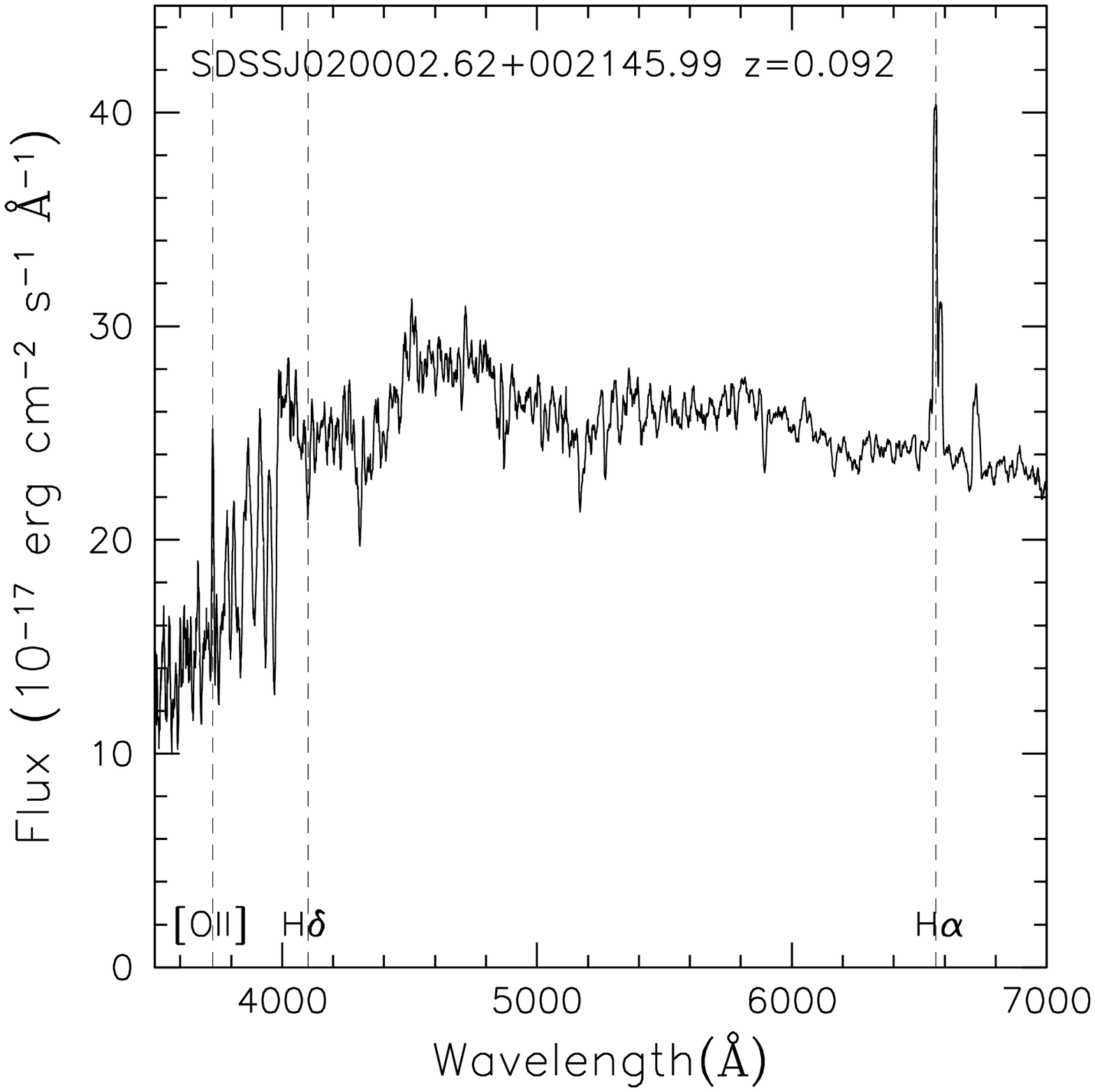}
}
\centering{
\includegraphics[scale=0.25]{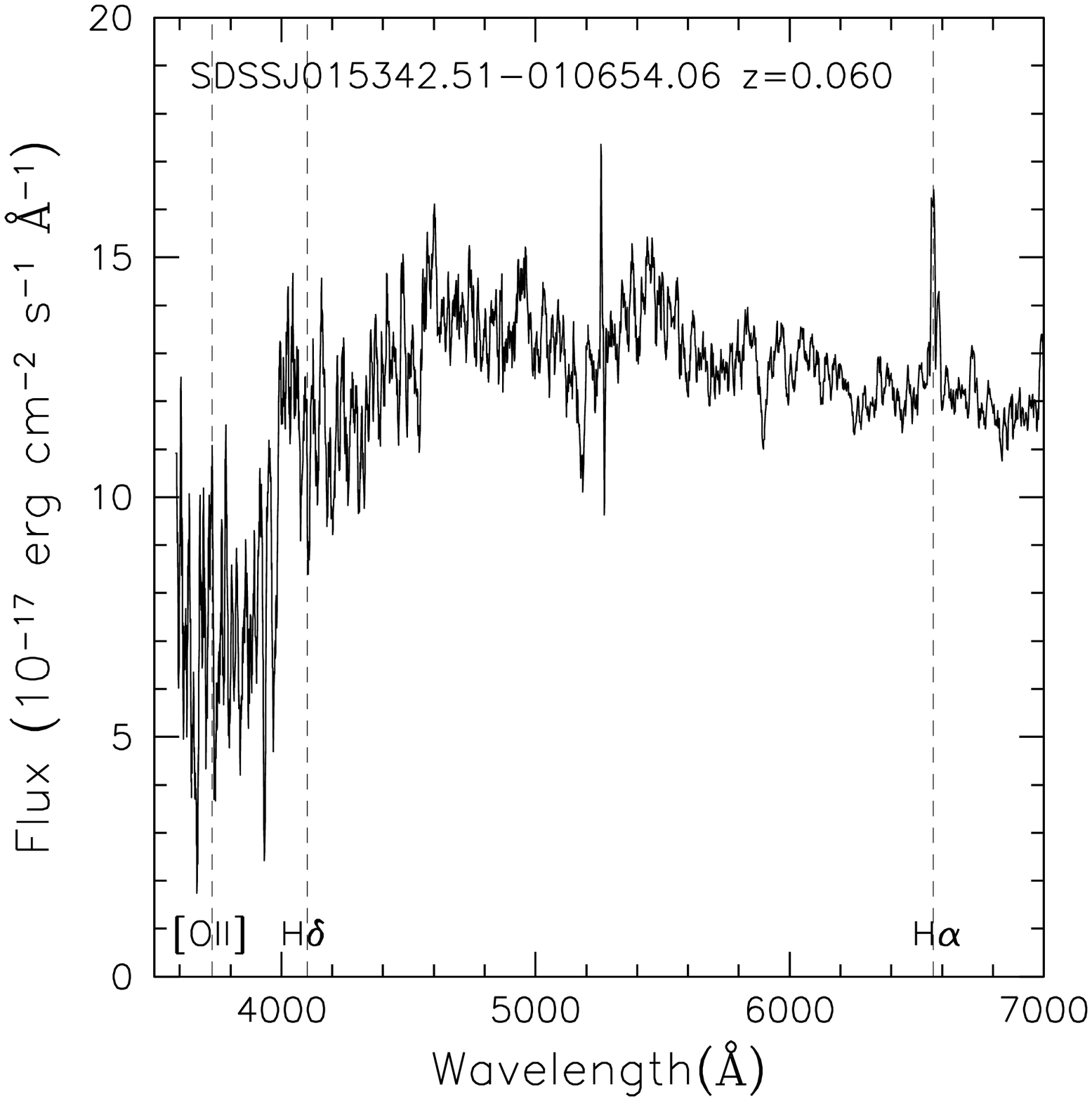}
\includegraphics[scale=0.25]{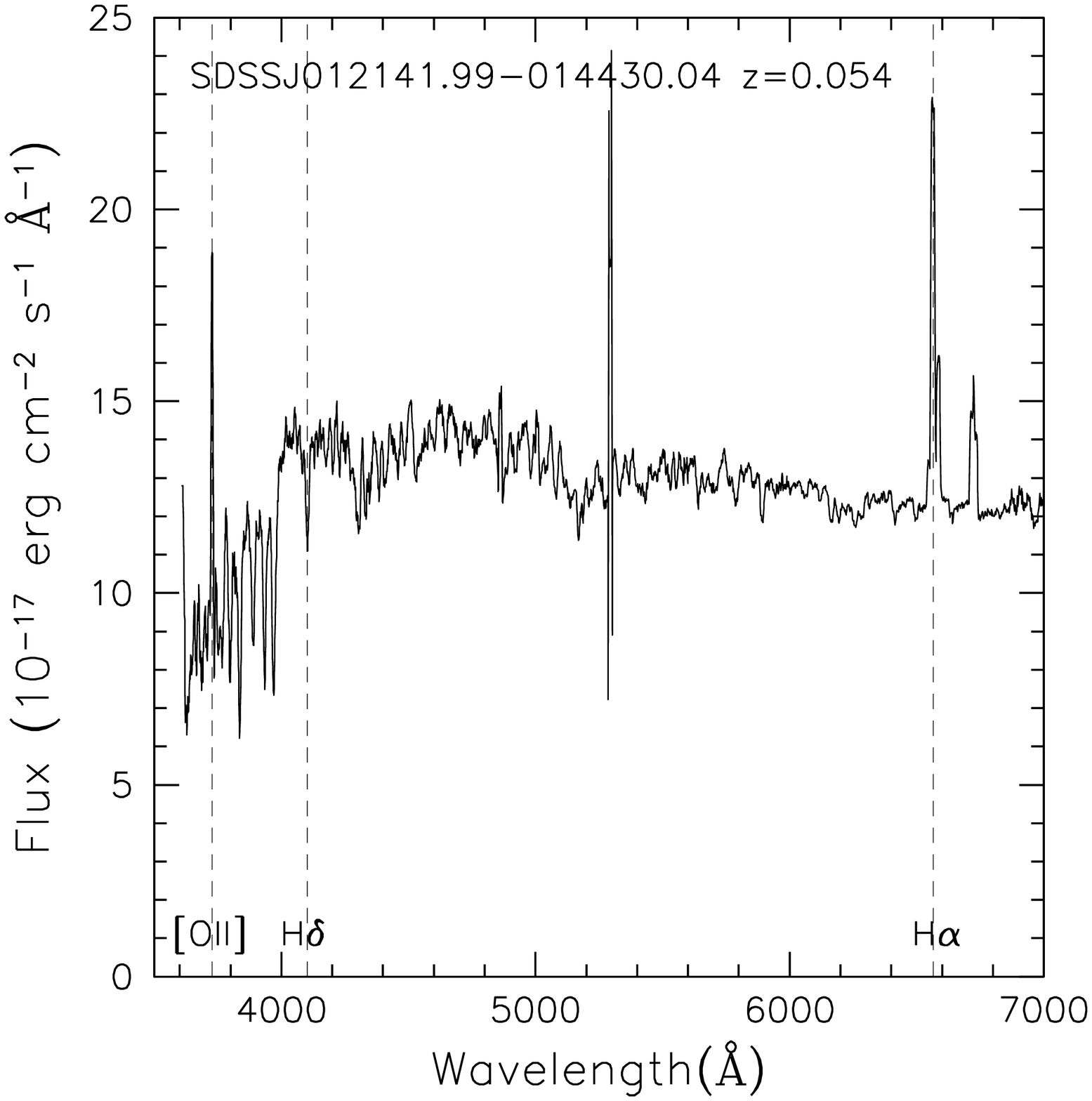}
\includegraphics[scale=0.25]{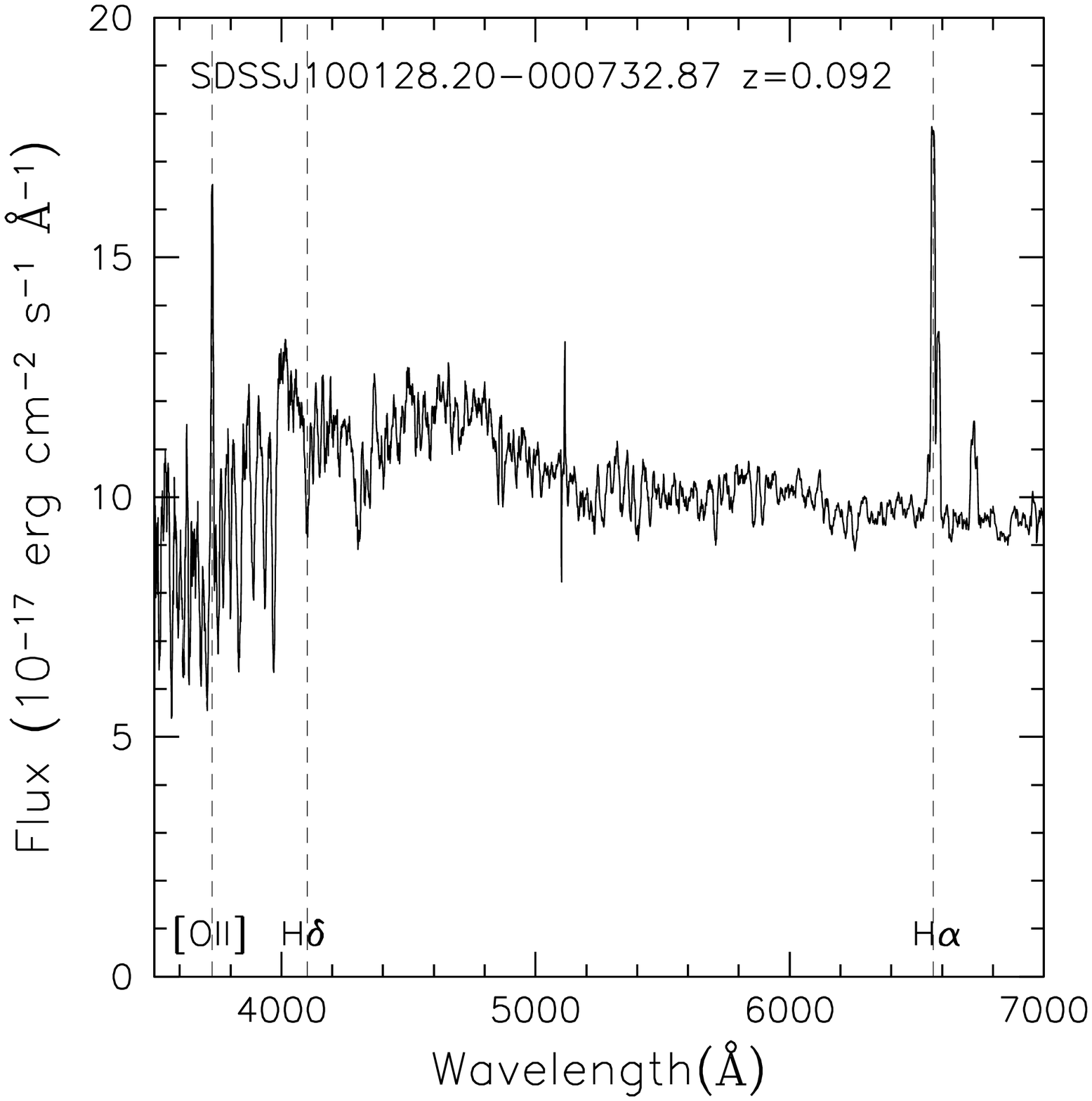}
}
\centering{
\includegraphics[scale=0.25]{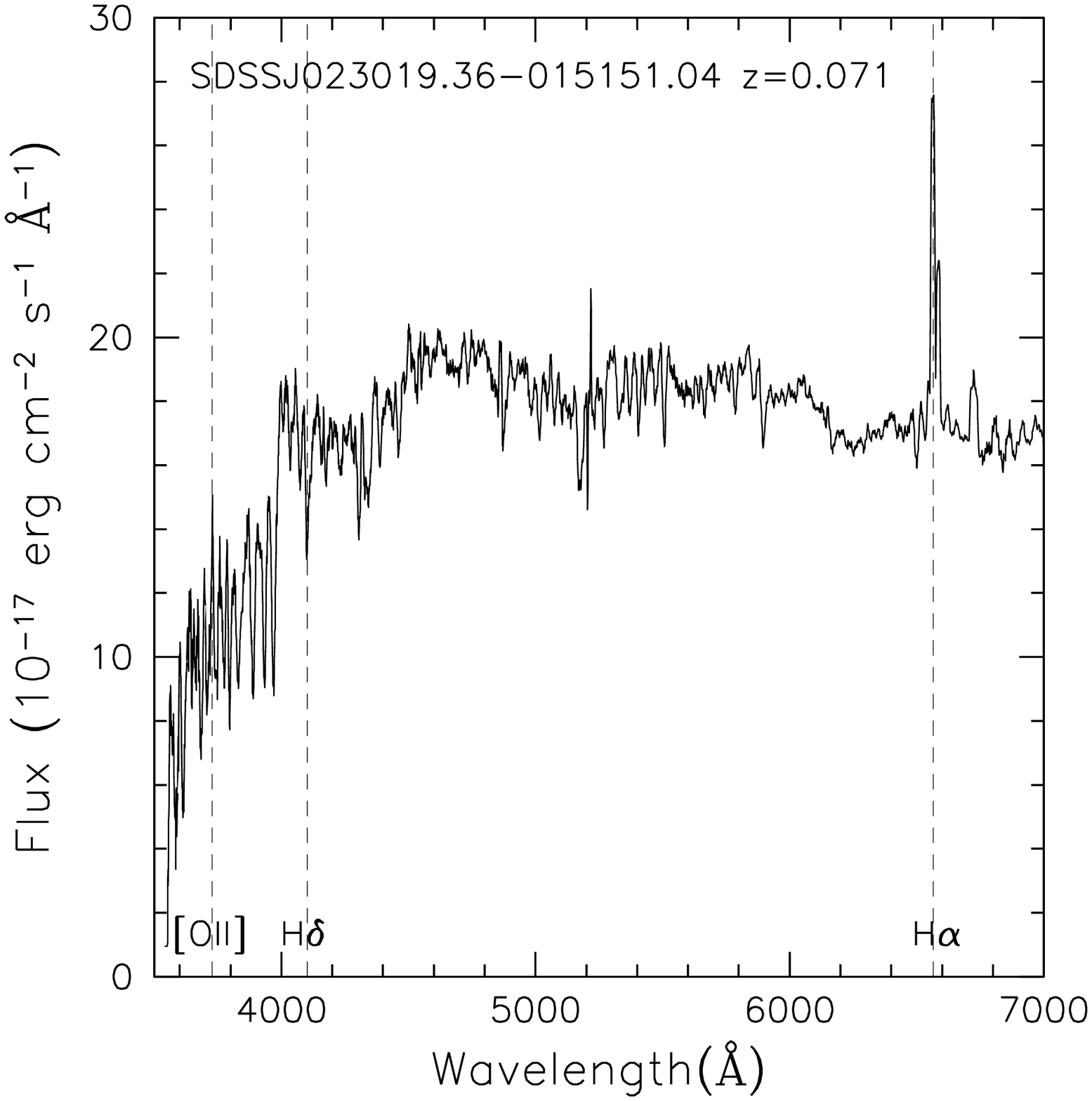}
\includegraphics[scale=0.25]{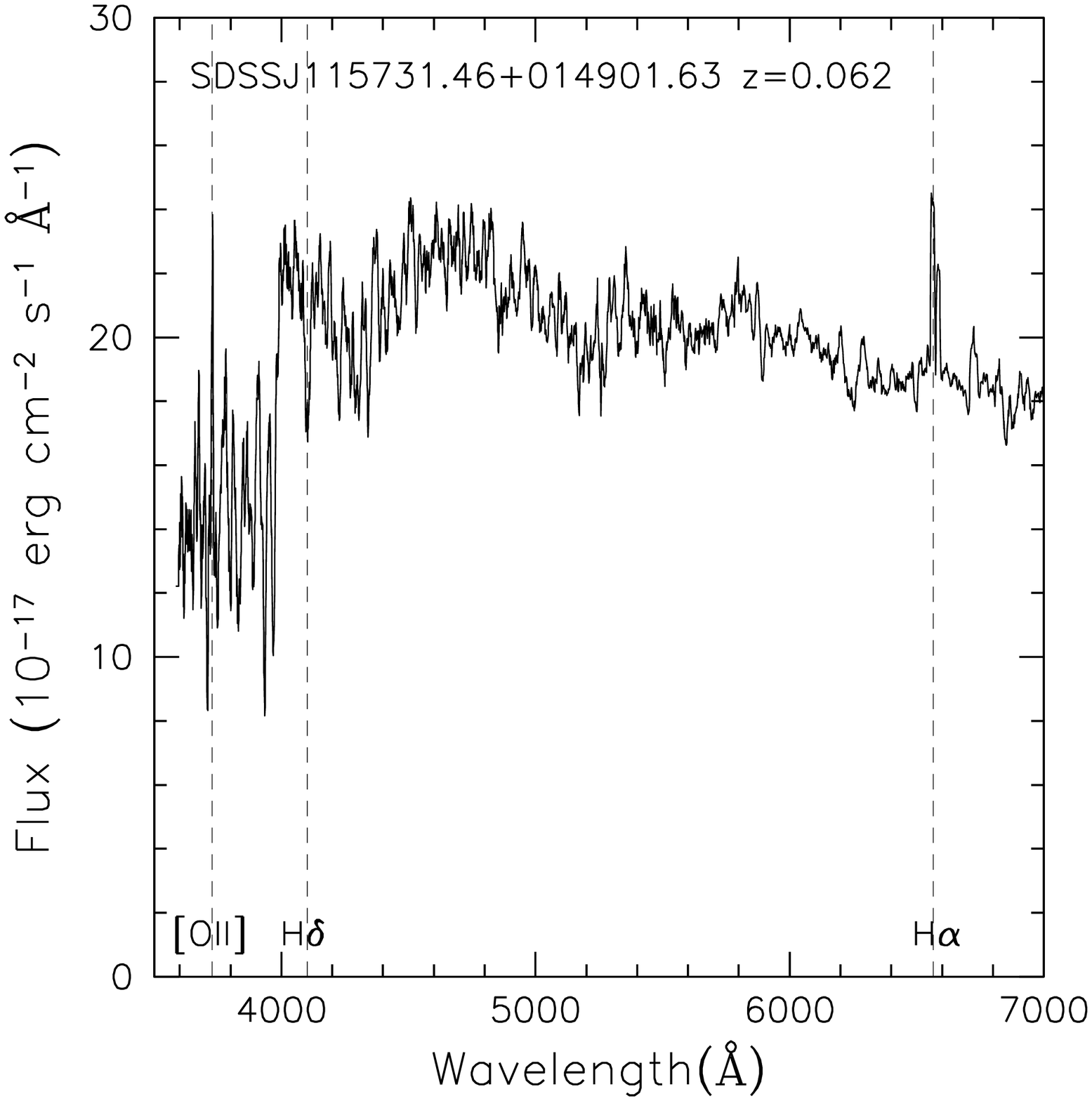}
\includegraphics[scale=0.25]{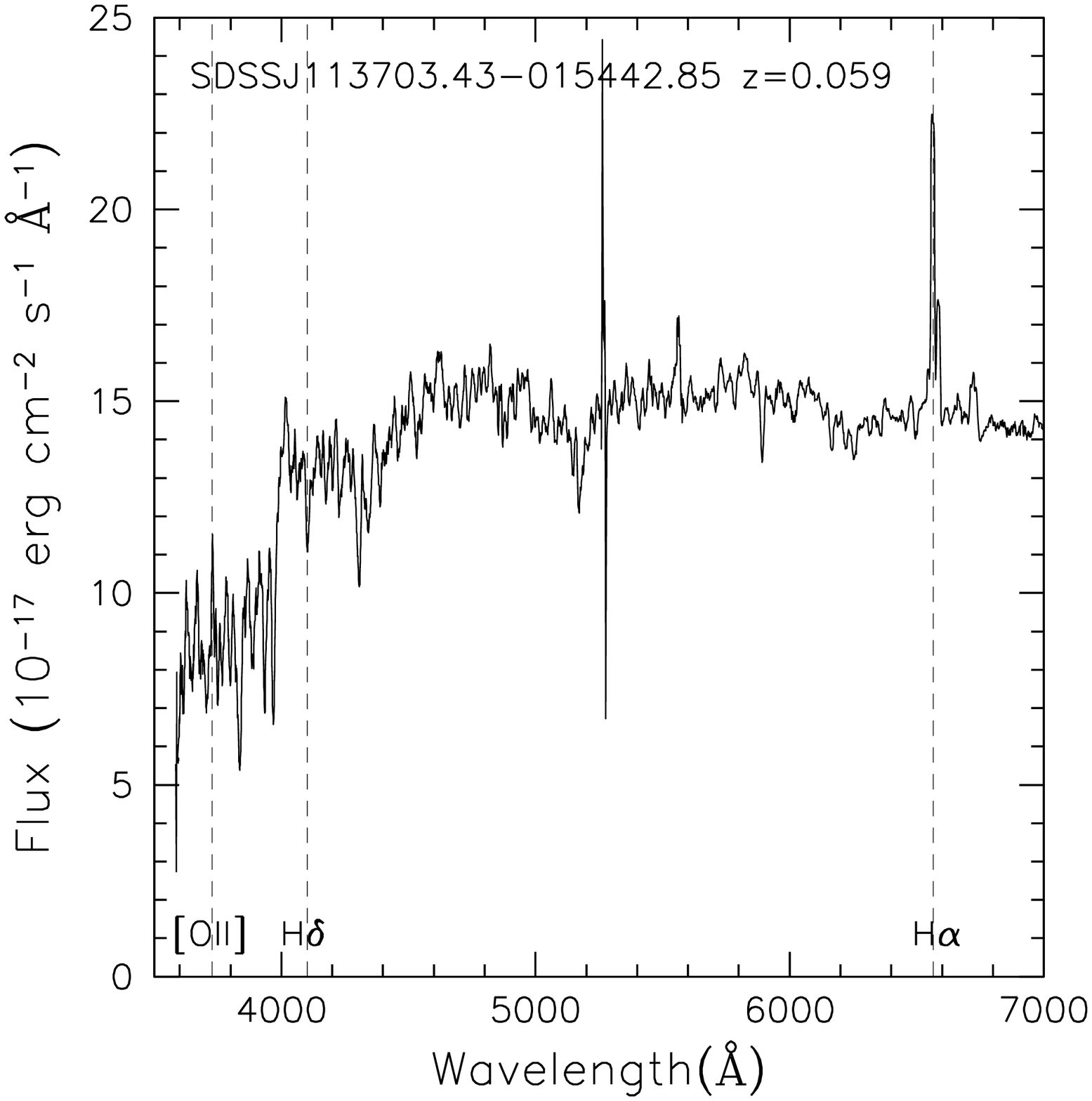}
}
\caption{
\label{fig:as_spectra}
 Example rest-frame spectra of active spiral galaxies. The spectra are
 shifted to the rest-frame and smoothed using a 10 \AA\ box.
 Each panel corresponds to
 that in figure \ref{fig:as_image}. 
}\end{figure}

  Figure \ref{fig:image} shows example images (30''$\times$30'') of
  passive spiral 
  galaxies. In figure \ref{fig:spectra}, corresponding spectra are
  shown. Unusual properties of these galaxies are already clear just by
  comparing these two figures, i.e., clear spiral arm structures are
  seen in the images, whereas there is no current star-formation activity
  as shown by the lack of [O{\sc ii}] and H$\alpha$ emission lines in the spectra. It
  is interesting to study from where these unique features originate. 
  For a comparison, we also select active (normal) spiral
  galaxies in our sample as galaxies with $C_{\rm in}>$0.5 and 1 $\sigma$
  detections in both [O{\sc ii}] and H$\alpha$ in emission. We removed
  galaxies with an AGN signature from the active spiral sample using a
  prescription given in Kewley et al. (2001) and  G{\' o}mez et al. (2003).
  When a galaxy satisfied all three line ratio criteria to be an AGN
  (Figure 15, or   equations 5, 6, and 7 of Kewley et al. 2001),
  we removed it from our  sample as an AGN.
 Images and spectra of active (normal) spiral galaxies are shown in 
   figures \ref{fig:as_image} and \ref{fig:as_spectra}. Compared with these
  galaxies, passive spirals have smoother profiles.

  Among 25813 galaxies in our volume-limited sample (0.05$<z<$0.1 and
  $Mr^*<-20.5$), there are 73 (0.28$\pm$0.03\%) passive spiral galaxies
  in total. The 
  number of active spirals is 1059 (4.10$\pm$0.12\%). The relatively small
  percentages stem from our stringent criteria for the inverse concentration
  parameter, $C_{\rm in}$.

\section{Environment of Passive Spiral Galaxies}

 \subsection{Local Galaxy Density}\label{density}

 First, we clarify the environment where passive spiral galaxies exist. 
 For each galaxy, we measure the
 projected metric distance to the 5th nearest galaxy (within $\pm$3000
 km s$^{-1}$ of
 the galaxy in redshift space) from within the same volume-limited sample as
 defined above (0.05$<z<$0.1 and $Mr^*<-20.5$). If the projected area
 (in ${\rm Mpc^2}$), enclosed by the 5th nearest neighbor 
 distance, touched the boundaries of the SDSS data, we corrected the area
 appropriately for the amount of missing area outside the survey
 boundaries. Then, we divided 5 (galaxies) by the area subtended by the
 5th nearest 
 galaxy to obtain the local galaxy density in Mpc$^{-2}$. 
   This methodology allowed us to quote pseudo
 three-dimensional local galaxy densities for all of our galaxies in the
 volume-limited sample, and did not require us to make any correction
 for background  and foreground contamination.

 In figure \ref{fig:density}, we plot the density distribution of passive
 spiral  
 galaxies by the dotted line.  The solid line shows the distribution of all
 galaxies in our volume-limited sample. The long-dashed line shows the
 distribution 
 of cluster galaxies defined as galaxies within 0.5 Mpc from the nearest
 C4 galaxy cluster (G{\' o}mez et al. 2003) in the angular
 direction and within
 $\pm3000\ {\rm km\, s^{-1}}$ from the redshift of a cluster. 
 Kolomogorov--Smirnov tests show that all of the distributions are
 different from each other with more than a 99.9\% significance
 level. The environment of 
 passive spiral galaxies is not in a cluster region, nor in the general
 field as a normal galaxy. 
 On the contrary, the distribution of passive spiral galaxies
 is right in the middle of cluster galaxies and field
 galaxies. The dotted line also shows that passive spiral galaxies avoid
 cluster core regions; at the 
 same time they do not show the same distribution as field galaxies.
 For a comparison, we plot a distribution of active (normal) spiral
 galaxies by the short-dashed line. Compared with that of all galaxies,
 the distribution
 slightly shifts to a less-dense environment, as expected from the
 morphology--density relation. The distribution is different from that of
 passive spirals with more than 99.9\% significance. 

\subsection{Cluster-Centric Radius}\label{radius}

 In figure \ref{fig:radius}, we plot the distribution of passive spiral
 galaxies as a function of the cluster-centric radius. Here, 
 the cluster-centric radius is measured as the projected distance to the
 nearest 
 cluster  within $\pm$3000 km s$^{-1}$ from the cluster redshift.  The cluster
 list is taken from Miller et al. in preparation, which measures a cluster
 center as the position of the brightest cluster galaxy.
  The 
 physical distance is normalized to the virial radius using the relation in
 Girardi et al. (1998).
 We divide distributions by that of all galaxies and then normalize them to
 unity for clarity. Note that comparisons of fractions between different
 curves are meaningless due to this normalization.
  The dotted, hashed line shows the distribution of passive spiral galaxies. 
 The solid lines show that of active (normal) spiral galaxies. The dashed line
 is for early-type galaxies selected using  $C_{\rm in}$ parameter
 ($C_{\rm in}<$0.33) 
 with no constrains on the emission lines. The
 fraction of early-type galaxies is higher in smaller
 cluster-centric radius and that of spiral galaxies are higher in
 larger-radius regions, which represents the so-called ``the
 morphology-density relation'' (Dressler 1980; Dressler et al. 1997;
 Postman, Geller 1984;  Fasano et al. 2000; Goto et al. 2003c). Passive  
 spiral galaxies are preferentially in 1--10 virial radius, which suggests
 that they exist in cluster infalling regions along with the previous figure.

\subsection{Photometric and Spectroscopic Properties}

 In figure \ref{fig:gri}, we plot the distribution of passive spiral
 galaxies in the rest-frame $g-r$ vs. $r-i$ plane. Instead of Petrosian
 magnitude, we  
 use the model magnitude to compute the colors of galaxies since
 the signal-to-noise  ratio is higher for the model magnitude. In the
 SDSS, the model magnitudes are 
 measured using the Petrosian radius measured in $r$ band. (Thus the same
 radius is used to measure the model magnitudes in 5 filters. See Stoughton
 et al. 2002 for more details on the SDSS magnitudes.) 
 The observed colors are $k$-corrected
 to the restframe using the $k$-correction given in Blanton et
 al. (2003b; v1\_11). The contours show the distribution of all galaxies in
 the volume-limited sample for
 comparison. A peak of the contour around ($g-r$, $r-i$) = (0.75, 0.4)
 consists of elliptical galaxies. The distribution of spiral galaxies
 extends to the bluer direction in both $g-r$ and $r-i$.
 Interestingly, passive spiral galaxies are almost as red as
 elliptical galaxies in $g-r$, reflecting the truly passive nature of these
 galaxies. Note that because colors are photometrically determined, they
 are free from the aperture bias. In $r-i$ color, some passive spirals
 are almost as blue as spiral galaxies. 
 Although a detailed comparison with population synthesis models is
 needed, a combination of dust extinction and low metallicity might be
 able to explain the bluer $r-i$ color.

 In figure \ref{fig:rk}, we present restframe $J-K$ vs. $r-K$ colors of
 passive and active galaxies by open and solid dots, respectively. 
 Infrared colors are obtained by matching our galaxies to the Two Micron
 All Sky Survey (2MASS; Jarrett et al. 2000) data. Infrared magnitudes
 are shifted to the rest-frame using $k$-corrections given in Mannucci et
 al. (2001). 
  If passive spiral galaxies are actually star-forming, but emission 
  lines are invisible at the optical wavelength due to heavy obscuration 
  by dust, then such dusty galaxies should have a redder color in $r-K$ by
 $\sim$1 mag (e.g., see Figure 2 of Smail et al. 1999).
 Among our sample
 galaxies, 31/73 passive
 spirals (9317/25813 all galaxies) were measured with 2MASS.  
 As in the previous figure, the solid lines show distribution of all
 galaxies in 
 the volume-limited sample.  Compared with the solid lines, active spirals
 show a slightly bluer distribution in $r-K$. 
 There are three passive spirals around $r-K\sim$1.5, which might have
 star formation hidden by dust at optical wavelength. 
 The majority of passive spirals, however,  do not show
 significantly bluer distribution in $r-K$ than all galaxies, suggesting
 that the majority of passive spirals do not have dust-enshrouded star
 formation.

 In figure \ref{fig:hd}, we plot the distributions of H$\delta$ EW for
 passive spirals (dotted lines), active spirals (dashed lines) and all
 galaxies (solid lines) in the volume-limited sample. H$\delta$ EWs are
 measured using the flux-summing method discussed in Goto et
 al. (2003b). The flux-summing method is robust for weak absorption
 lines and noisy spectra. In the figure, the absorption lines have positive
 EWs. As shown in the figure, passive spirals have very weak H$\delta$
 absorption peaked at around 0 \AA.
 In contrast, active spirals have much stronger H$\delta$
 absorption. Since H$\delta$ absorption becomes strong only when A-type stars
 are dominant in the galaxy spectra, it is a good indicator of
 the post-starburst phase of a galaxy. For example, E+A galaxies (Zabludoff
 et al. 1996; Balogh et al. 1999; Dressler et al. 1999; Poggianti et
 al. 1999) are thought 
 to be post-starburst galaxies since they do not have any current star
 formation (no [O{\sc ii}] nor H$\alpha$ in emission), but do have many A-type
 stars (strong H$\delta$ absorption). 
 Such a phase can only appear when
 a starbursting galaxy truncates its starburst (a post-starburst
 phase). 
 Therefore, small H$\delta$ EWs of passive spirals
 indicate that they are not in a post-starburst phase. The origin of
 passive spirals are likely to be different from that of E+A (or
 post-starburst) galaxies, which recently have been found to be
 merger/interaction related (Goto et al. 2003b,d).
 Passive spirals seem to have stopped their star formation
 gradually, rather than experiencing sudden truncation.

\section{Discussion}\label{discussion}

 In subsection \ref{selection}, we have selected an unusual population of
 galaxies with a spiral morphology and without emission lines such as
 H$\alpha$ and [O{\sc ii}]. The optical color--color diagram
 (figure \ref{fig:gri}) also revealed that these galaxies are as red as
 elliptical galaxies, reflecting the passive nature of these galaxies. 
 One possible explanation for these galaxies is heavy obscuration by
 dust. In such a case, passive spiral galaxies might have star-formation
 activity just as normal galaxies, but the star-formation might be hidden by dust.
  The scenario could be consistent with both of our observational results:
 lack of emission lines and red colors in the optical wavelength. 
 However, in $r-K$ color (figure \ref{fig:rk}), the majority of passive spiral
 galaxies do not appear to be much redder than normal galaxies. 
 For example, radio-detected
 galaxies in Smail et al. (1999) are redder than other galaxies by
 $\sim$1 mag in $r-H$.
  Therefore, figure \ref{fig:rk} is against the dust-enshrouded scenario, which
 should results in a very  
 red $r-K$ color, and thus suggests the truly passive nature of these
 galaxies. In addition, it is not likely that very dusty galaxies
 preferentially live in cluster infalling regions.

 In subsection \ref{density}, we reveal that passive spiral galaxies
 preferentially exist in cluster infalling regions, using both the local
 galaxy density (figure \ref{fig:density}) and the cluster-centric radius
 (figure \ref{fig:radius}). This is direct evidence to connect
 the origins of these galaxies to the cluster environment. The
 characteristic environments are 1--2 Mpc$^{-2}$ in the local galaxy
 density and 1--10 virial radius in the cluster-centric radius. Quite
 interestingly, these environments coincide with the characteristic density
 and radius where star-formation rate declines toward the cluster center
 or the 
 dense environment.  G{\' o}mez et al. (2003) and Lewis et al. (2002) studied
 the star-formation rate in a galaxy as a function of the cluster-centric radius
 and the local galaxy density, and found that the star-formation rate declines
 around the same environment as we found in the present
 study. Furthermore, Goto et al. (2003c) studied the morphology--density
 relation using similar SDSS data (0.05$<z<$0.1 and $Mr^*<-20.5$)
 and an automated galaxy classification (Yamauchi et al. in preparation).
 They found that the morphological 
 fraction of galaxies starts to change approximately at the same
 environment as our study: the fraction of S0 and elliptical galaxies
 starts to increase (and Sc galaxies decrease) toward cluster center, or
 larger galaxy density right  at around 1 cluster-centric virial radius, or local galaxy density 1 Mpc$^{-2}$.        
      These coincidences in the environment suggest that the same mechanism might be
 responsible for all of the effects happening: the creation of passive
 spiral galaxies; the decrease of the galaxy star-formation rate; and the 
 morphological change in the relative galaxy fraction. These coincidences
 might be explained naturally by the following interpretation:
  As galaxies
 approach this critical environment (1--2 Mpc$^{-2}$ or 1--10
 virial radius), they stop their star formation as 
 shown in  G{\' o}mez et al. (2003), by changing spiral galaxies into passive
 spiral galaxies as found in this study. If a spiral galaxy stops star
 formation calmly without its morphology being disturbed, it is likely to
 develop to be a S0 galaxy (Bertin, Romeo 1988; Bekki et al. 2002) as is seen in the
 morphology--density relation of Goto et 
 al. (2003c). According to this scenario, passive spirals are likely to
 be a population of galaxies in transition. 
      In addition, there have been many results from other observations
 and surveys that supports 
 this scenario. Abraham et al. (1996) reported that cluster members
 become progressively bluer as a function of the cluster-centric distance
 out to 5 Mpc in Abell 2390 ($z=0.23$). 
  Terlevich, Caldwell, and Bower (2001) reported that $U-V$
 colors of early-type galaxies are systematically bluer outside of the
 core of the Coma cluster.  
   Pimbblet et al. (2002) studied 11 X-ray luminous clusters
 (0.07$<z<$0.16), and found that median galaxy color shifts bluewards
 with decreasing local  galaxy density. 
   At higher redshift, Kodama et al. (2001) reported that the colors of
 galaxies abruptly change at sub-clump regions surrounding a cluster
 at $z=0.41$. Although it is difficult to directly compare this
 environment with ours due to the different definitions of the local galaxy
 density, it is highly possible that their color change occurs in  the
 same environment that we found. 
  van Dokkum et al. (1998) found S0 galaxies in the outskirt of a
 cluster at $z=0.33$. These S0s show a much wider scatter in their
 colors, 
 and are bluer on average than those in cluster cores, providing
 possible evidence for recent infall of galaxies from the field. In
 addition, many studies reported that star formation in the cores of
 clusters is much lower than that in the surrounding field (e.g., Balogh
 et al. 1997, 1998, 1999, 2002; Poggianti et al. 1999; Martin et al. 2000;
 Couch et al. 2001).

 The existence of passive spiral
 galaxies also brings us a hint about the origin of these three phenomena.
 It supports a transformation of galaxies, which does not
 disturb the arm structures of spiral galaxies. Possible preferred
 candidate mechanisms include ram-pressure stripping (Spitzer, Baade
 1951; Gunn, Gott 1972; Farouki, Shapiro 
 1980; Kent 1981; Fujita 1998; Abadi et al. 1999; Fujita, Nagashima 1999;
 Quilis et al. 2000) and simple removal of gas
 reservoir (Larson et al. 1980; Balogh et al. 1999; Bekki et al. 2001,2002). 
 The evaporation of the cold gas in disc galaxies via heat conduction
 from the surrounding hot ICM might be able to 
 reduce the star-formation rate in a galaxy (Cowie, Songaila 1977; Fujita 2003).
 It has been known that preheating of the intergalactic medium
 can affect the morphologies of galaxies by strangling the gas accretion
 (strangulation; Mo, Mao 2002; Oh, Benson 2002). In fact, Finoguenov et al. (2003a)
 found filamentary gas in the Coma cluster, and predicted the existence
 of passive spirals around the filament. 
  Although the characteristic environment (1--2 Mpc$^{-2}$ or 1--10
 virial radius) might seem to have slightly too low density
 for ram-pressure or strangulation to occur, it is possible for galaxy
 sub-clumps around a cluster to have local hot gas that is dense enough for
 stripping (Fujita et al. 2003). Indeed, Kodama et al. (2001) found that
 galaxy colors change  at such sub-clumps around a cluster. 

 Perhaps major merger/interaction origins are less preferred since
 such dynamical processes disturb the arm structures in spiral galaxies, and
 thus do not result in creating passive spirals. For example, Finoguenov
 et al. (2003b) found a disturbed signature in M86, proposing that  M86
 is likely to be merger originated.
 The weak H$\delta$
 absorption lines shown in figure \ref{fig:hd} also support a quiescent
 transformation of galaxies. 
 However, we can not
 exclude a minor merger origin since such a process might be able to
 occur without disturbing the spiral arms. 
 In their morphology--density relation study, Goto et al. (2003c)
 observed a decrease of S0s and an increase of 
 ellipticals at cluster cores (virial radius $<$ 0.3 or galaxy density
 $>$ 6 Mpc$^{-2}$), and 
 proposed that the major merger/interaction might be
 dominant in the cluster core regions.
 The proposal is consistent with
 our results, which showed devoid of passive spiral galaxies within 0.6
 virial radius, 
 or greater than $\sim$3 Mpc$^{-2}$ in local galaxy density. 
   On the other hand, some theoretical work predicts that 
 it is difficult to have frequent merger/interaction in cluster
 cores since the relative velocities of galaxies are so high in such regions
 (Ostriker 1980; Binney, Tremaine 1987; Mamon 1992; Makino, Hut 1997).   
 In such a case, S0s (or passive spirals) might simply fade away to be a
 small  elliptical galaxy. 
  In summary, implication for the  cluster core regions is 
 either (i) passive spiral galaxies merged into large elliptical
 galaxies in cluster cores, or (ii) the discs of passive spiral galaxies completely fade
 away to become small elliptical galaxies.

 Also, in terms of cluster galaxy evolution, passive spiral galaxies might fit
 well with the previous observational results. It has been known that a
 fraction of blue galaxies is larger in higher redshift (the
 Butcher--Oemler effect; Butcher, Oemler 1978, 1984; Couch, Sharples
 1987; Rakos, Schombert 1995; Couch et al. 1998; Margoniner, de Carvalho
 2000; Margoniner et al. 2001; Ellingson 
 et al. 2001; Kodama, Bower 2001; Goto et al. 2003a) and 
 that the fraction of cluster spiral galaxies are also larger in the
 past (Dressler et al. 1997;  Couch et al. 1998; Fasano et al. 2000;
 Diaferio et al. 2001; Goto et al. 2003a). Many people have speculated
 about the
 morphological transformation from spiral galaxies to S0 galaxies
 (e.g., Dressler et al. 1997; Smail et al. 1998; Poggianti et al. 1999;
 Fabricant et al. 2000; Kodama, Smail 2001).
   Fraction of early-type galaxies in rich clusters are smaller in the
 past (Andreon et al. 1997; Dressler et al. 1997; Lubin et
 al. 1998; van Dokkum et al. 2000).  
   From a morphological point of view, since passive spiral galaxies have
 already stopped star 
 formation, in the near future, their disc structures  will become fainter and 
 fainter, to be seen as disc galaxies with smoother profile, i.e., possibly
 S0 galaxies. Spectrally, passive spirals are already almost as red as elliptical
 galaxies, but their spiral arms must have experienced star-formation activity
 until recently; therefore, a passive spiral galaxy itself must have been much
 bluer in the past, just like the blue population of galaxies that are numerous in
 the higher redshift clusters. Therefore, although this is not direct
 evidence, it is very likely that passive spiral galaxies are a
 population of galaxies in transition, during the course of the Butcher--Oemler
 effect and the morphological Butcher--Oemler effect. 
     Perhaps, it is also worth noting that E+A (K+A or post-starburst)
 galaxies often thought to be cluster-related
 are found to have their origin in merger/interaction with accompanying
 galaxies (Goto et al. 2003b,d), and thus E+A galaxies are not likely to be
 a product of the morphological transition in the cluster regions.

\section{Conclusions}\label{conclusion}
 
 Using a volume-limited sample of the SDSS data, we studied the
 environment of passive spiral galaxies as a function of the local galaxy
 density and cluster-centric radius. Since passive spirals were only
 found in cluster regions in previous work, this was the first attempt to
 select passive spirals uniformly, in all environments. It is found that passive
 spiral galaxies exist in a local galaxy density of 1--2 Mpc$^{-2}$ and
 1--10 virial radius. Thus, the origins of passive spiral galaxies
 are likely to be cluster-related. These characteristic environments
 coincide with the environment where the galaxy star-formation rate suddenly
 declines (Lewis et al. 2002;  G{\' o}mez et al. 2003) and the fractions of
 galaxy morphology start to deviate from the field value (Goto et
 al. 2003c). Therefore, it is likely that the same physical mechanism is
 responsible for all of these observational results: the
 morphology-density relation; the decline of star formation rate; and the
 creation of passive spiral galaxies. The existence of passive spiral
 galaxies suggest that a physical mechanism that works calmly is
 preferred to dynamical origins such as a major merger/interaction since such a
 mechanism can destroy the spiral-arm structures. 
  Passive spiral galaxies are likely to be a galaxy population in
 transition between red, elliptical/S0 galaxies in low redshift clusters
 and blue, spiral galaxies numerous in higher redshift clusters as seen
 in the Butcher--Oemler effect and the morphological Butcher--Oemler
 effect. 

\bigskip


 We are grateful to Sidney van den Bergh, Don York, Masayuki Tanaka,
 Robert C. Nichol, Andrew Hopkins,  Alex Finoguenov and Michael Balogh
 for valuable comments, which 
 contributed to improve the paper.  
 We thank an anonymous referee for useful comments.
 T.G. acknowledges financial support from the Japan Society for the
 Promotion of Science (JSPS) through JSPS Research Fellowships for Young
 Scientists. 

 Funding for the creation and distribution of the SDSS Archive has been
 provided by the Alfred P. Sloan Foundation, the Participating
 Institutions, the National Aeronautics and Space Administration, the
 National Science Foundation, the U.S. Department of Energy, the
 Japanese Ministry of Education, Culture, Sports, Science and
 Technology, and the Max Planck Society. The SDSS Web site is
 $\langle$http://www.sdss.org/$\rangle$. 

 The SDSS is managed by the Astrophysical Research Consortium (ARC) for the Participating Institutions. The Participating Institutions are The University of Chicago, Fermilab, the Institute for Advanced Study, the Japan Participation Group, The Johns Hopkins University, Los Alamos National Laboratory, the Max-Planck-Institute for Astronomy (MPIA), the Max-Planck-Institute for Astrophysics (MPA), New Mexico State University, Princeton University, the United States Naval Observatory, and the University of Washington.

\appendix

\section*{Aperture Bias}\label{aperture}

\begin{figure}
\begin{center}
\includegraphics[scale=0.7]{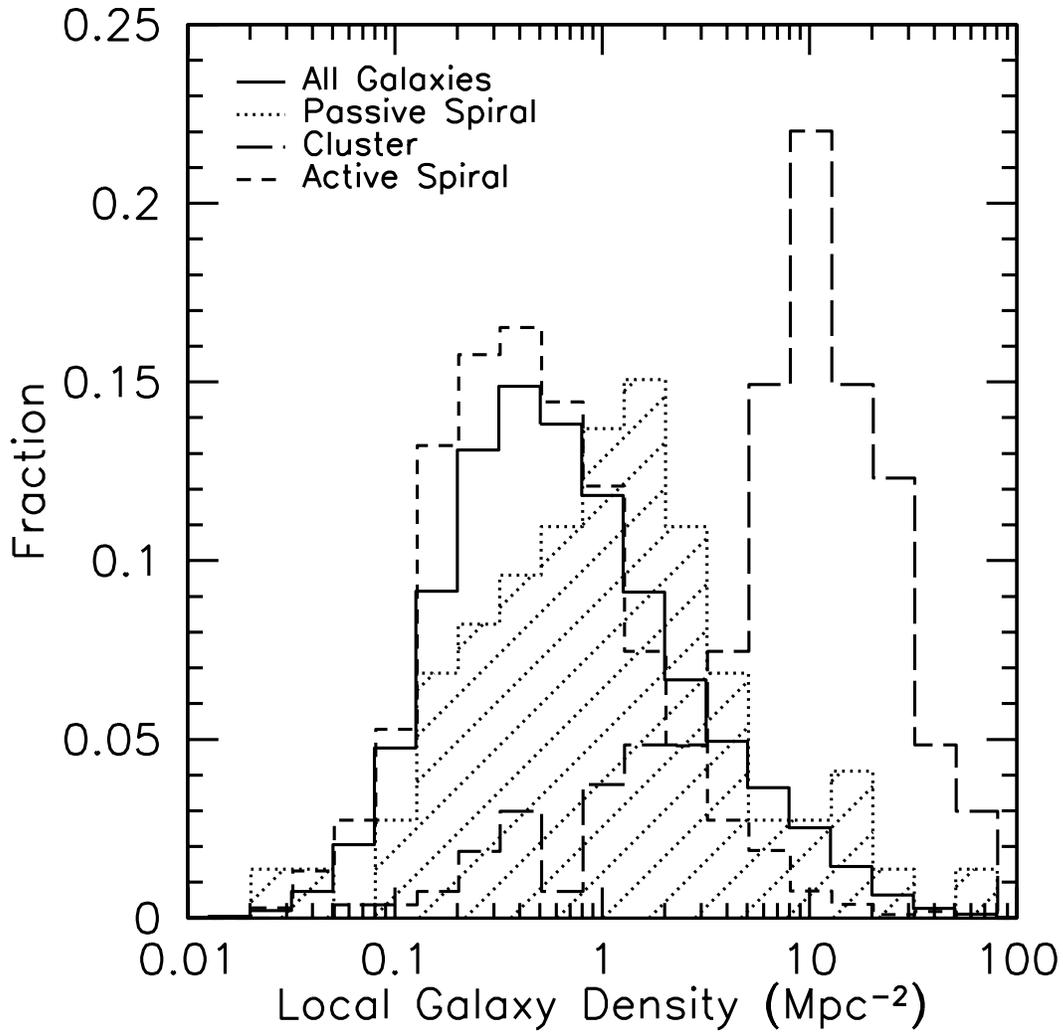}
\end{center}
\caption{
\label{fig:density}
 Distribution of the densities for passive spiral galaxies (hashed
 region) and all galaxies (solid line) in
a volume-limited sample.  A Kolomogorov--Smirnov test shows that distributions
 of passive spirals and
all galaxies are from a different distribution. The long dashed line shows
 the distribution of cluster galaxies. The short dashed line shows that of
 active spiral galaxies. }
\end{figure}

\begin{figure}
\begin{center}
\includegraphics[scale=0.7]{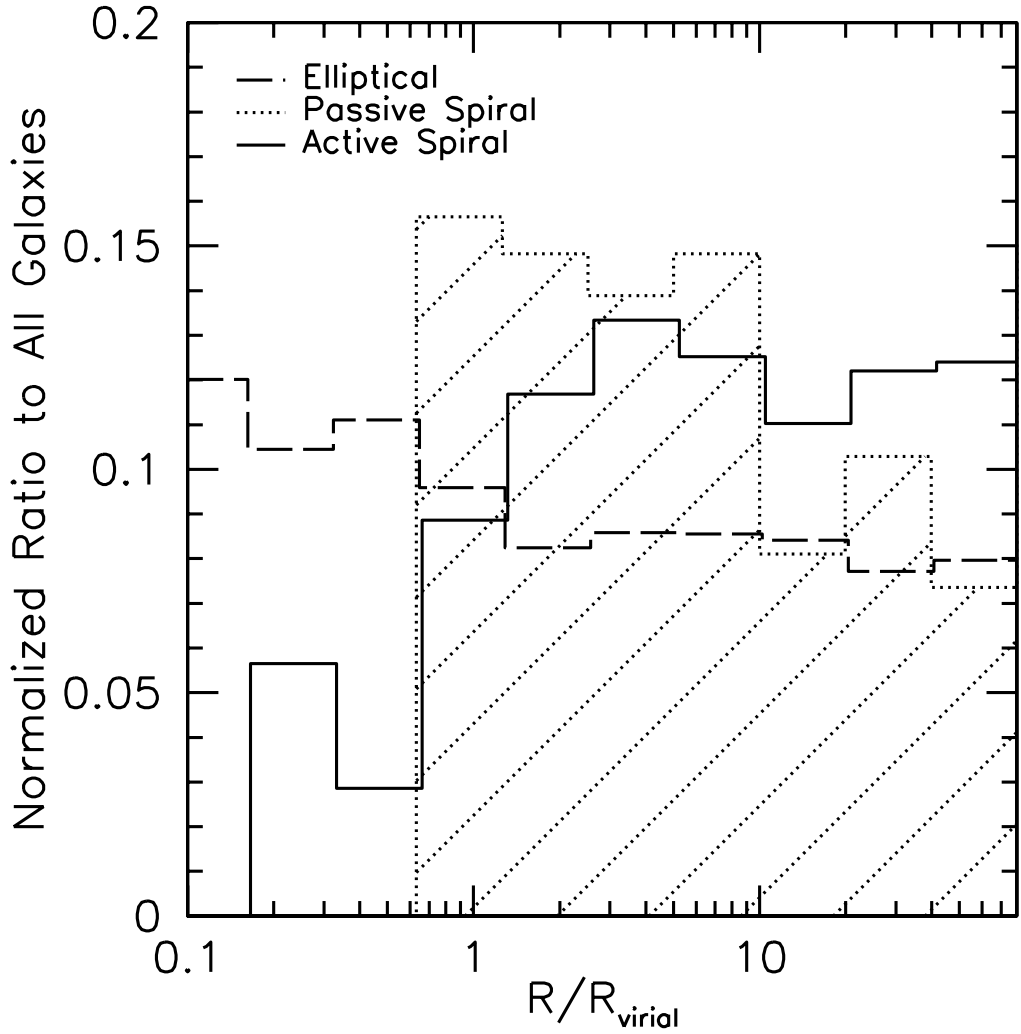}
\end{center}
\caption{
\label{fig:radius}
 Distribution of passive spiral galaxies as a function of
 cluster-centric radius. The dotted, dashed, and solid lines show the
 distributions of passive spiral, elliptical and active spiral galaxies,
 respectively. The distributions are relative to that of all galaxies in
 the volume-limited sample and normalized to be 1 for clarity.
 The cluster-centric radius is measured as a distance to a
 nearest C4 cluster (Miller et al. in preparation) within $\pm$3000 km
 s$^{-1}$, and normalized by the  virial radius (Girardi et al. 1998). 
  }
\end{figure}

\begin{figure}
\begin{center}
\includegraphics[scale=0.7]{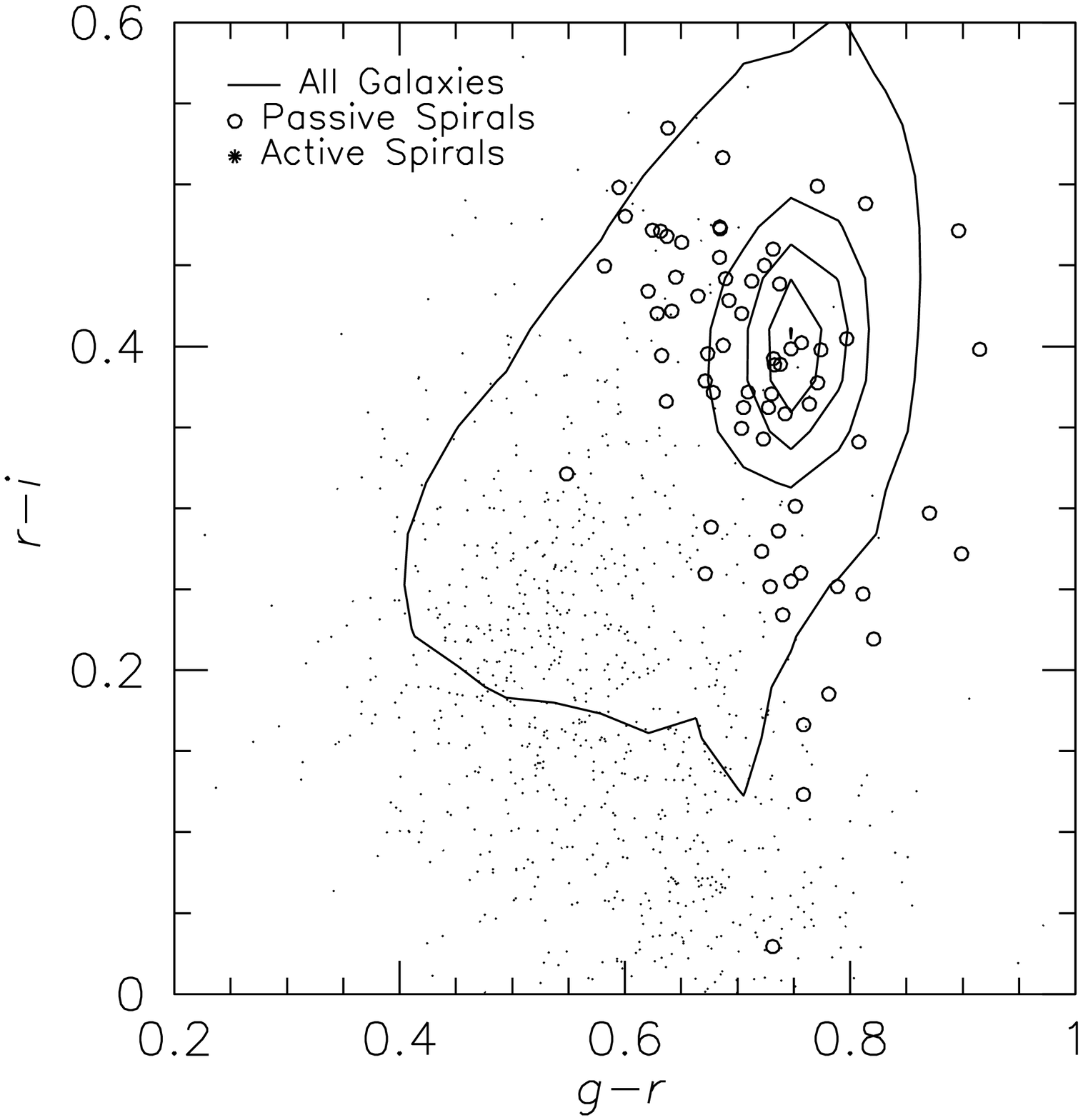}
\end{center}
\caption{
\label{fig:gri}
 Distribution of passive spirals in rest-frame $g-r$ vs. $r-i$ 
 plane. The contours show the distribution of all galaxies in our volume-limited sample. 
 The open circles and filled dots represent passive and
 active spiral galaxies, respectively.
  }
\end{figure}

\begin{figure}
\begin{center}
\includegraphics[scale=0.7]{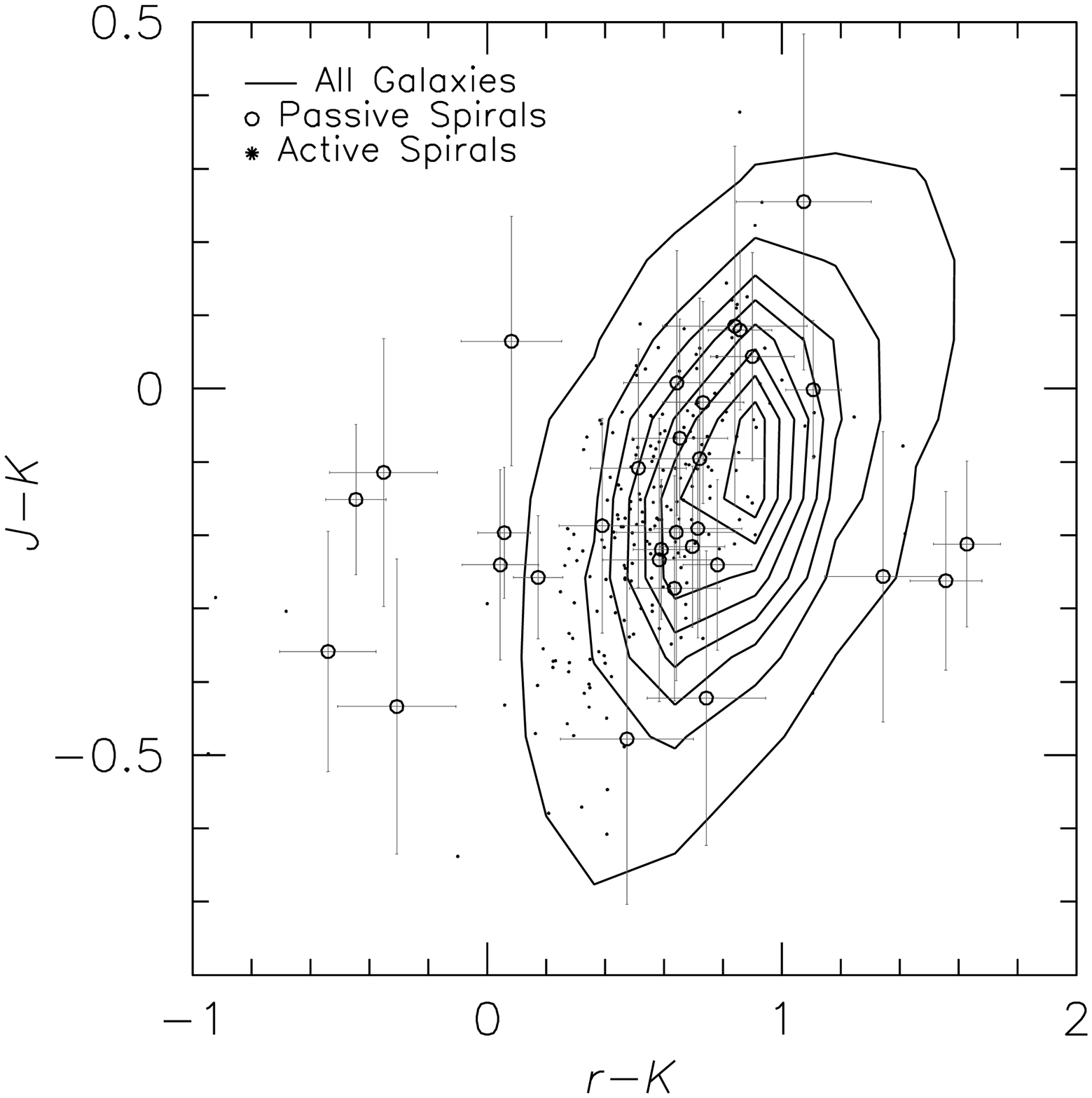}
\end{center}
\caption{
\label{fig:rk}
 Distribution of passive spirals in restframe $J-K$ vs. $r-K$
 plane. The contours show the distribution of all galaxies in our volume-limited sample. 
 The open circle and filled dots represent passive and
 active (normal) spiral galaxies, respectively.
  }
\end{figure}

\begin{figure}
\begin{center}
\includegraphics[scale=0.7]{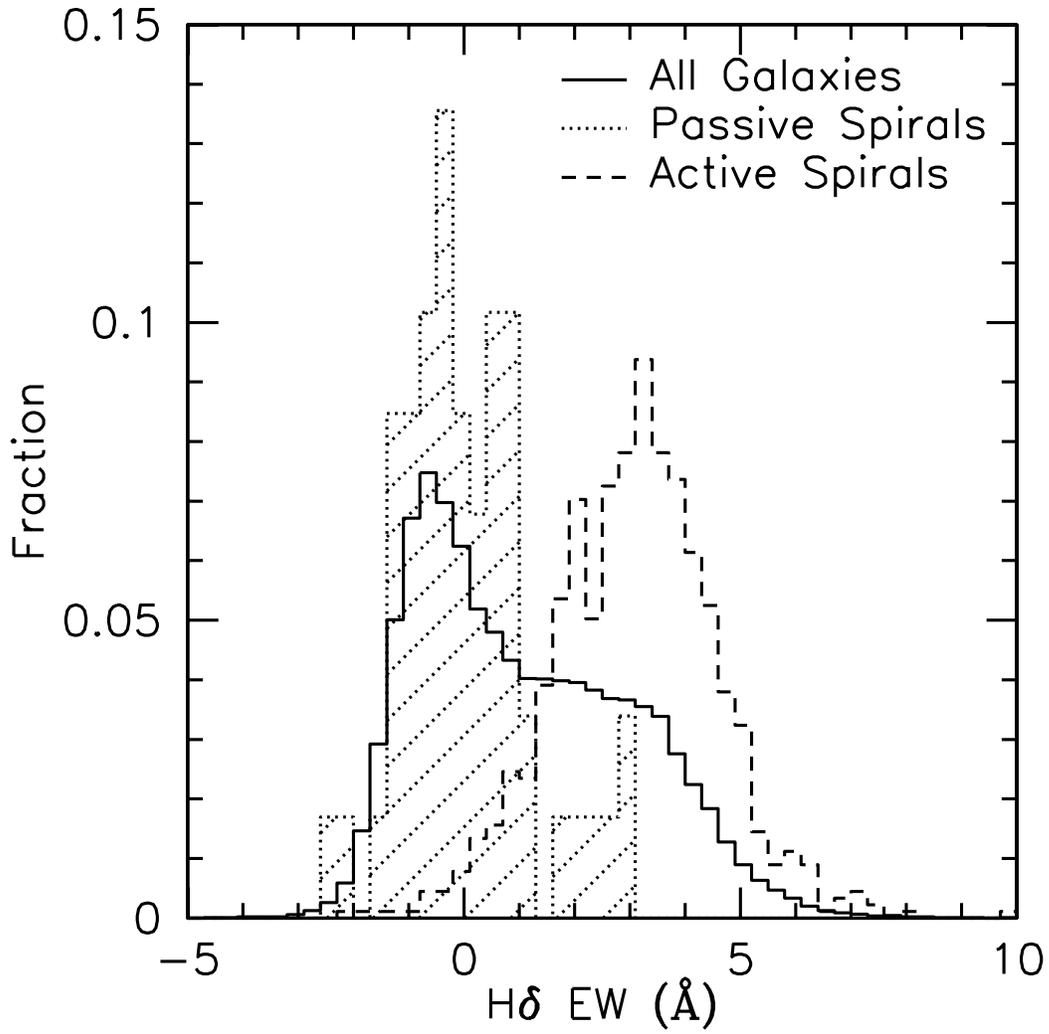}
\end{center}
\caption{
\label{fig:hd}
 Distributions of H$\delta$ EWs of passive spiral galaxies, active
 spiral galaxies, and all
 galaxies in the volume-limited sample. The solid, dashed and dotted lines
 are for all galaxies, active spiral galaxies,
 and passive spiral galaxies, respectively. The absorption lines are
 positive in this figure.  Passive spiral galaxies tend
 to have weak H$\delta$ absorption lines.
  }
\end{figure}

\begin{figure}
\begin{center}
\includegraphics[scale=0.7]{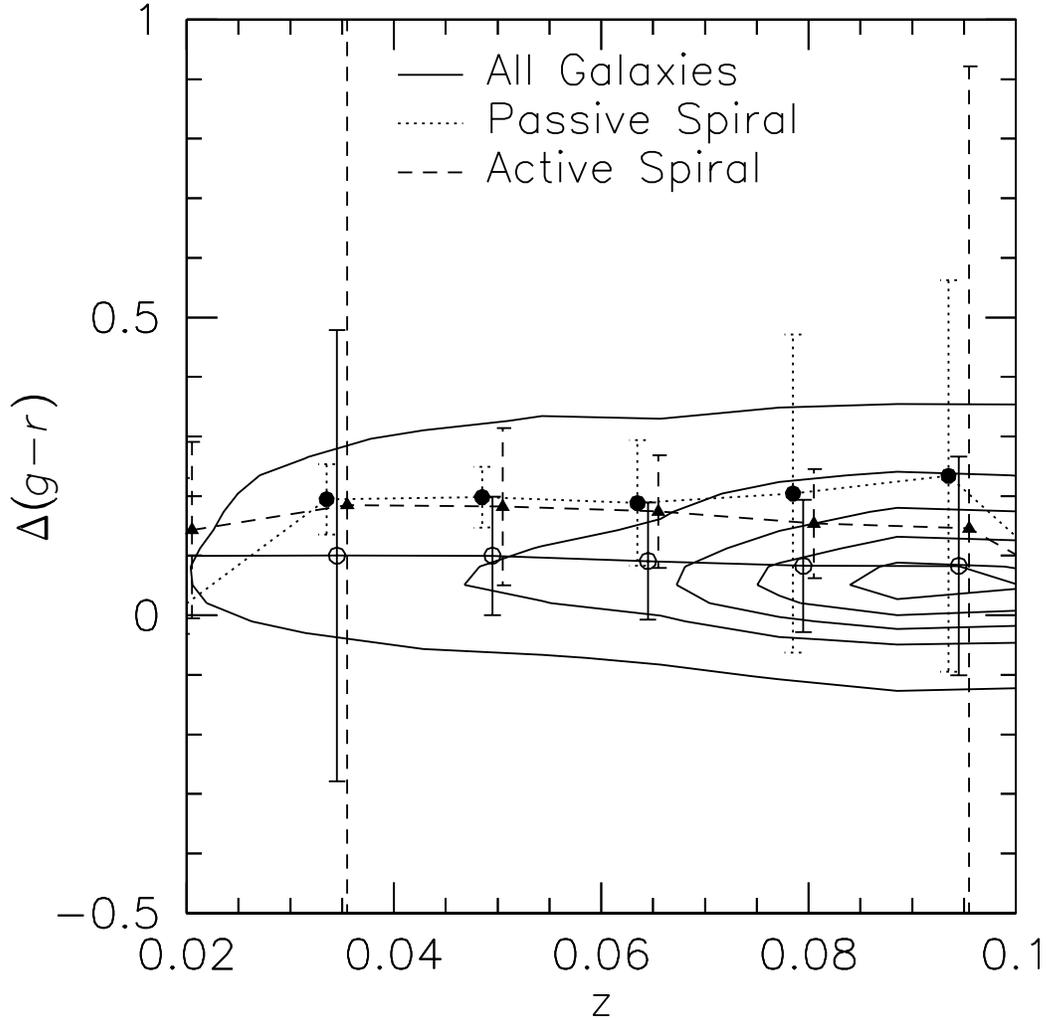}
\end{center}
\caption{
 Differences between the fiber color (within 3'' aperture) and model color
 (using Petrosian radius measured in $r$) plotted against
 redshift. The solid, dotted, and dashed lines show the medians of all galaxies,
 passive spirals, and active spirals, respectively. 
 The difference, $\Delta(g-r)$, should be smaller
 at a higher redshift since 3'' fiber can collect a larger amount of total galaxy
 light at higher redshift.
 Both passive and active spirals have larger
 $\Delta(g-r)$ than all galaxies since they are less concentrated. 
 Throughout the redshift range we used (0.05$<z<$0.1), $\Delta(g-r)$ of
 passive spirals is consistent with a constant within the error,
 suggesting that the aperture effect is not severe within the redshift
 range that we used.
  }\label{fig:color_gradient}
\end{figure}

\begin{figure}
\begin{center}
\includegraphics[scale=0.7]{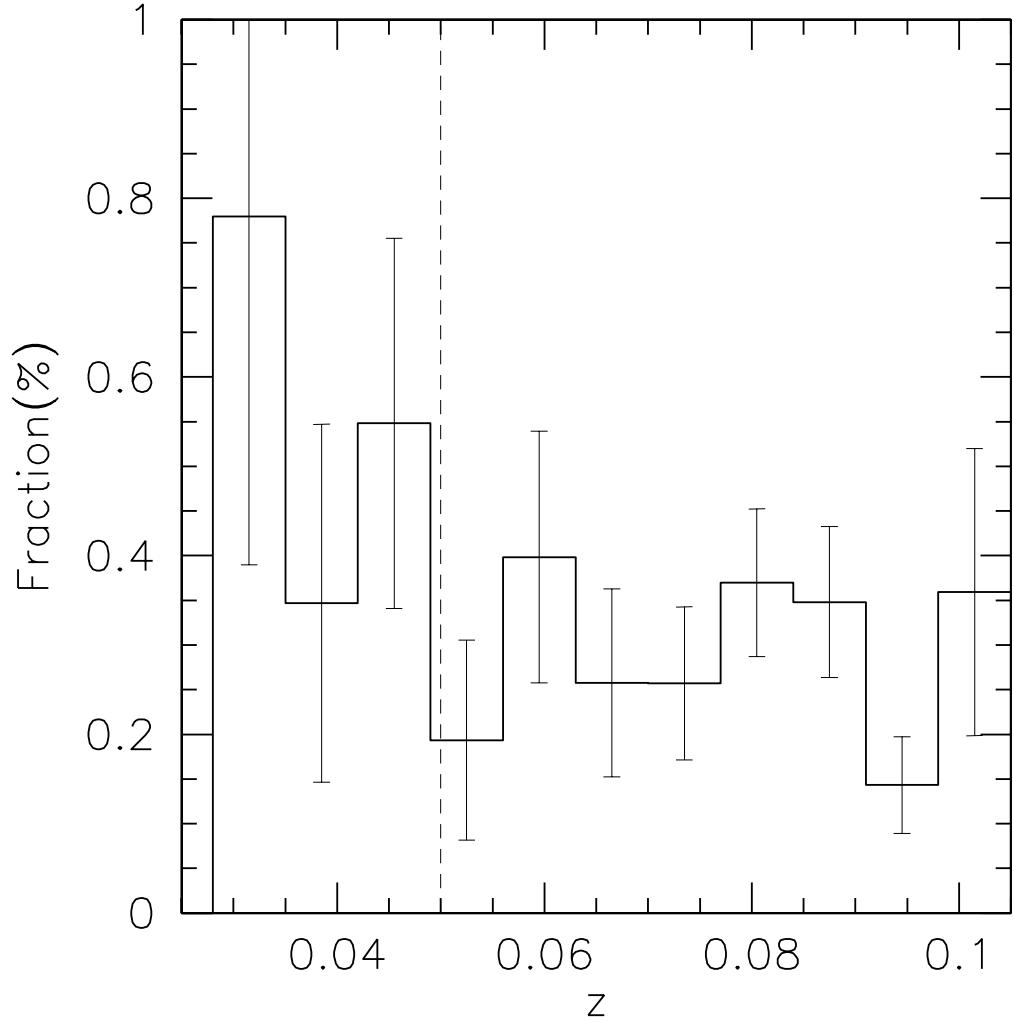}
\end{center}
\caption{
 Fractions of passive spiral galaxies (in percentage) to all galaxies
 among the volume-limited sample
 shown as a function of redshift. Our sample includes passive spiral
 galaxies only between $z=0.05$ and $z=0.1$, where fractions are consistent
 to be constant, suggesting that the aperture bias is not a strong effect in our
 sample.  
  }\label{fig:aperture}
\end{figure}

 Since SDSS spectroscopy is performed with a fiber spectrograph, which
 captures light within a 3 arcsec aperture, aperture bias is a
 concern. 
  Aperture bias could result in an increase of passive spiral
 galaxies with decreasing redshift since at a lower redshift, a 3 arcsec
 fiber misses more light from a disc of a galaxy. Using the data from
 LCRS with a 3.5 arcsec fiber spectrograph, Zaritsky et al. (1995)
 showed that at $z>0.05$, the spectral classifications of galaxies are
 statistically unaffected by aperture bias. Using a similar sample of
 the SDSS galaxies,  G{\' o}mez et al. (2003) also
 limited their galaxies to $z>0.05$, and proved that aperture bias does
 not change their results. We followed these two authors and 
 limited our sample with $z>$0.05 to minimize this potential bias.
 In the main analysis of the paper, there are several pieces of evidence
 suggesting that these passive spiral galaxies are not seriously biased
 by the aperture effect. In figure \ref{fig:gri}, passive spiral
 galaxies are much redder than normal galaxies. To calculate colors, we
 used a model magnitude 
 which uses the Petrosian radius in $r$ for all colors (Stoughton et
 al. 2002), and is thus free from 3'' aperture bias.  
 Therefore, the red colors of these galaxies suggest that they are truly
 passive systems, and not an artifact of the aperture effect. 
 Also, in figure \ref{fig:density}, we compare the density distribution of
 passive spirals with normal star-forming spirals. The two distributions
 are statistically different. Again, if passive spirals are the artifact
 of aperture bias, the density distributions of star-forming and passive
 spiral galaxies should be similar. Therefore, this difference suggests
 that passive nature of these galaxies is truly unique to them. 
 Figure \ref{fig:color_gradient} shows a difference in $g-r$ color between
  the fiber magnitude (measured with 3'' aperture) and the model magnitude
 (measured using Petrosian radius in $r$, usually larger than 3'',
 especially in low redshift) as a
 function of redshift. The color difference, $\Delta (g-r)$, is defined as follows.
  \begin{equation} 
  \Delta (g-r) = (g-r)_{\rm fiber}- (g-r)_{\rm Petrosian}.
   \end{equation}
 Therefore, $\Delta (g-r)$ is larger for more extended nearby galaxies
 with a
 strong color gradient.
 In figure \ref{fig:color_gradient}, the contours show the distribution
 of all galaxies in 
 our volume-limited sample. The solid, dotted and dashed lines show the medians
 of all galaxies, passive spirals, and active spirals. Since both passive
 and active spirals are less concentrated, their medians have somewhat
 higher values than all galaxies. If aperture bias is severe,
 $\Delta(g-r)$ should be much larger at a lower redshift since the
 difference between a 3'' aperture and the Petrosian radius of galaxies is
 larger. However, in figure \ref{fig:color_gradient}, $\Delta(g-r)$ of
 passive spirals is
 almost constant throughout the redshift range that we used
 (0.05$<z<$0.1). 
  The figure suggests that the aperture effect is constant within the
 redshift range, and thus does not invite a strong redshift dependent bias.
  In figure \ref{fig:aperture}, we present the fraction of passive spiral galaxies as a
 function of redshift. It clearly shows a strong aperture effect at
 $z<0.05$. However, throughout this paper, we limit our sample between z=0.05 and
 0.1, where fractions of passive spirals are constant
 within the error. This suggests that the aperture bias is not strong within
 our sample. 
   Since the aperture effect results in missing the light from outer
 disc of spiral galaxies, another possibility is that if the
 distribution of  spiral galaxies with a large bulge depend on the
 environment, we might have an environment dependent bias. However, we know
 that spiral galaxies with a large bulge are numerous in a dense environment
 such as cluster cores (Dressler et al. 1997; Fasano et al. 2000; Goto
 et al. 2003c). Nevertheless, we did not find passive spirals in cluster
 core regions (Figures \ref{fig:density}, \ref{fig:radius}), suggesting that
 the passive spiral galaxies are not the product of an environmental
 dependence of large bulge spirals plus the aperture bias.
   We end this section by quoting that Hopkins et al. (2003)
 compared the star-formation rate estimated from H$\alpha$ (SDSS data;
 subject to 3 arcsec aperture
 bias) and that from the radio flux (FIRST data; i.e., with no aperture
 bias), concluding that both star-formation rate estimates agree with
 each other after correcting the H$\alpha$ flux using the ratio of the 3'' fiber
 magnitude to the Petrosian (total) magnitude in $r$ band.

\clearpage

\clearpage

\clearpage

\clearpage

\clearpage

\clearpage

\clearpage

\clearpage

\clearpage

\clearpage

\clearpage

\clearpage

\begin{table}[h]
\begin{center}
\caption{
Wavelength ranges used to measure [O{\sc ii}] EW, H$\alpha$ EW, and  H$\delta$ EW. 
}\label{tab:wavelength}
\begin{tabular}{llll}
\hline
  & Blue continuum  & Line & Red continuum \\
\hline
\hline
$[{\rm OII}]$            &  3653--3713 \AA &    3713--3741 \AA & 3741--3801 \AA  \\ 
H$\alpha$          &  6490--6537 \AA &    6555--6575 \AA & 6594--6640 \AA  \\ 
H$\delta$          &  4030--4082 \AA &    4088--4116(4082-4122) \AA &
 4122--4170 \AA  \\ 
\hline
\end{tabular}
\end{center}
\end{table}

\end{document}